\documentclass[12pt,preprint]{aastex}

\begin{document}
\def\la{\mathrel{\hbox{\rlap{\hbox{\lower4pt\hbox{$\sim$}}}\hbox{$<$}}}}
\def\ga{\mathrel{\hbox{\rlap{\hbox{\lower4pt\hbox{$\sim$}}}\hbox{$>$}}}}
\def\lam{$\lambda$}
\def\kms{km~s$^{-1}$}
\def\vphot{$v_{\rm phot}$}
\def\ang{~\AA}
\def\syn{SYNOW}
\def\dm15{{$\Delta$}$m_{15}$}
\def\rsi{$R$(Si~II)}
\def\v10{$V_{10}$(Si~II)}
\def\wsi{$W_lambda$(Si~II)}
\def\vdot{\.v(Si~II)}
\def\W575{$W(5750)$}
\def\W610{$W(6100)$}
\def\6100{the 6100~\AA\ absorption}
\def\tex{$T_{\rm exc}$}
\def\ve{$v_{\rm e}$}

\title {Comparative Direct Analysis of Type~Ia Supernova Spectra.
III. Premaximum}

\author {David Branch\altaffilmark{1}, M.~A. Troxel\altaffilmark{1},
David~J. Jeffery\altaffilmark{1}, Kazuhito Hatano\altaffilmark{2},
Miriam Musco\altaffilmark{3}, Jerod Parrent\altaffilmark{1},
E.~Baron\altaffilmark{1}, Leeann Chau
Dang\altaffilmark{1,}\altaffilmark{4}, D.~Casebeer\altaffilmark{1},
Nicholas Hall\altaffilmark{1}, \& Wesley Ketchum\altaffilmark{1}}

\altaffiltext{1} {Homer L. Dodge Department of Physics and Astronomy,
University of Oklahoma, Norman,~OK 73019; e-mail: branch@nhn.ou.edu}

\altaffiltext{2} {Division of Network Business, PionetSoft Corporation,
  Shibuya-ku, Tokyo 151-0072, Japan}

\altaffiltext{3} {Department of Astronomy, Indiana University,
  Bloomington,~IN 47405}

\altaffiltext{4} {Department of Astronomy, Whitman College, Walla
  Walla,~WA 99362}

\begin{abstract}

A comparative study of spectra of 21 Type~Ia supernovae (SNe~Ia)
obtained about one week before maximum light, and 8 spectra obtained
11 or more days before maximum, is presented.  To a large extent the
premaximum spectra exhibit the defining characteristics of the four
groups defined in Paper~II (core-normal, broad--line, cool, and
shallow--silicon).  Comparisons with SYNOW synthetic spectra show that
all strong features and most weak ones can be accounted for in a
plausible way.  The issues of detached high--velocity features, the
possible ubiquity of carbon clumps, the maximum detectable ejecta
velocities, and the possibility of blueshifted emission--line peaks
are discussed.

\end{abstract}


\section{INTRODUCTION}

This is the third in a series of papers on a comparative direct
analysis of the optical spectra of Type~Ia supernovae (SNe~Ia).
Paper~I (Branch et~al. 2005) was concerned with a time series of
spectra, from 12 days before to 115 days after the time of maximum
light, of the well--observed, spectroscopically normal, Type~Ia
SN~1994D.  Paper~II (Branch et~al. 2006) concentrated on
near--maximum--light spectra of 24 SNe~Ia.  In this paper we focus on
pre--maximum--light spectra, which form in the outermost layers of the
ejected matter and provide clues to the nature of the explosions that
cannot be obtained from later spectra that form deeper in the ejecta.
Our emphasis is on line identifications, which are needed in the quest
to develop good hydrodynamical explosion models for SNe~Ia.  When
detailed spectrum calculations are carried out for explosion models
(e.g., Baron et~al. 2006; Howell et~al. 2001), discrepancies between
synthetic and observed spectra invariably appear.  Knowing the
identities of the observed features provides clues to how the
explosion models should be modified in order to achieve better
agreement with observation.

In Paper~II we divided the maximum--light spectra into four groups:
{\sl core--normal; broad--line; cool;} and {\sl shallow--silicon}
(denoted CN, BL, CL, and SS, respectively, in the figures and tables
of this paper).  The group assignments were made on the basis of
measurements of the (pseudo) equivalent widths of absorption features
near 5750\ang\ and 6100\ang, as well as on the appearance (depth,
width, and shape) of the 6100\ang\ absorption, which is produced by
the Si~II \lam6355 transition.  Although we framed the presentation
and discussion in terms of the four groups, in the end we concluded
that for the most part the spectra appeared to have a continuous
distribution of properties, rather than consisting of discrete
subtypes.  Therefore, exactly how the group boundaries were drawn was
not important.

Figure~1, a plot of the equivalent width $W(5750)$ of the absorption
feature near 5750\ang\ against the equivalent width \W610\ of the
absorption feature near 6100\ang\ in maximum--light spectra, is an
update of the corresponding figure in Paper~II, with the following
additional events: the core--normal SN~2003cg (Elias--Rosa
et~al. 2006), SN~2003du (Gerardy et al. 2004), and SN~2004S
(Krisciunas et~al. 2006); the broad--line SN~1983G (McCall
et~al. 1984) and SN~2002er (Kotak et al. 2006); and the
shallow--silicon SN~1999aa (Garavini et~al. 2004), SN~1999ac (Garavini
et~al. 2005), SN~2000E (Valentini et~al. 2003), SN~2005cg (Quimby
et~al. 2006), and SN~2003fg [the proposed super--Chandrasekhar SN~Ia,
also known as SNLS-03D3bb; Howell et~al. (2006); Jeffery, Branch, \&
Baron (2007a); but see also Hillebrandt, Sim, \& R\"opke (2007)].
SN~1999ac is a borderline case: in Figure~1 it falls between two
core--normals but on the basis of the shape, depth, and width of
\6100, we include it with the shallow--silicons.

To a large extent the premaximum spectra turn out to exhibit the
defining characteristics of the four maximum--light groups of
Paper~II, so we use the same terminology in this paper (with all group
assignments based on maximum--light, not premaximum, spectra).  Again,
we are not suggesting that SNe~Ia actually consist of discrete
subtypes; instead, we are thinking in terms of three inhomogeneous
groups radiating outward (in Figure~1) from the highly homogeneous
core normals.

In Figure~1, three lines of constant $W(5750)/W(6100)$ ratio are
shown.  This ratio is similar to although not the same as the well
known $R(Si~II)$ ratio (Nugent et~al. 1995), which is based on the
depths of the same two absorption features.  Considerable diversity at
a given $W(5750)/W(6100)$ ratio is evident, especially at the low
value of 0.05, e.g., the SN~2005hk, SN~1999ee, and
SN~2002bf have similar $W(5750)/W(6100)$ ratios but otherwise thei
spectra are very different.

In this paper, as in Papers~I and II, we confine our attention to
optical spectra, from the Ca~II H\&K feature in the blue
($\sim3700$\ang) to the Ca~II infrared triplet (Ca~II IR3) in the red
($\sim9000$\ang).  All spectra have been corrected for the redshifts
of the parent galaxies.  The original spectra have a range of slopes,
owing to intrinsic differences, differences in the amount of reddening
by interstellar dust, and observational error.  Because we are
interested only in the spectral features, and not in the underlying
continuum slopes, all spectra in this paper, both observed and
synthetic, have been flattened according to the local normalization
technique of Jeffery et~al. (2007b).  The flattening facilitates
comparison of the spectral features.  Mild smoothing also has been
applied to some of the spectra.

We examined all of the SN~Ia premaximum spectra available to us and
selected two samples: a ``one--week premax'' sample, consisting of one
spectrum each of 21 SNe~Ia observed between day~$-5$ and day~$-9$ with
respect to the time of maximum brightness in the $B$ band, and an
``early'' sample, consisting of one spectrum each of eight SNe~Ia
observed on or earlier than day~$-11$.  (All SNe of the early sample
are also in the one--week premax sample.)  The SNe~Ia and the epochs
of the selected spectra are listed in Table~1.

Continuing our attempt to provide an internally consistent
quantification of SN~Ia spectra, we have used the \syn\ code to fit
all spectra of Table~1.  Descriptions of \syn\ and its use can be
found in Branch et~al. (2003) and Paper~I\footnote{For consistency
with Papers~I and II, in this paper we do not use \syn\ version~2.0,
described and used by Parrent et~al. (2007).}.  Compared to other
supernova synthetic--spectrum codes, SYNOW is simple, but when working
in an empirical spirit from the data, as opposed to evaluating an
explosion model, SYNOW offers the advantage that the user has direct
control over line optical depths as function of radial coordinate,
while other codes input composition and density structures and then
compute the temperature structure and the line optical depths.  Thus
SYNOW is the most useful code for making line identifications (pending
eventual confirmation by means of detailed calculations for explosion
models whose spectra and light curves match the observations).  For
the present work, we adopt precepts for fitting spectra that are much
like those of Paper~II, e.g., the default excitation temperature is
7000~K for the cool SN~1986G and SN~1999by and 10,000~K for all
others.

\section{CORE--NORMALS}

Figure~2 shows the spectra of the seven core--normals of the one--week
premax sample.  The spectrum of SN~2001el differs from the others; as
shown by Mattila et~al. (2005), it resembles the day~$-14$ spectrum of
SN~1990N, so it will be discussed below, with the early sample.  The
other spectra show considerable homogeneity, although the diversity is
somewhat greater than in the maximum--light core--normal sample.  The
6100\ang\ absorption is deeper in SN~1998bu and SN~2003cg than in the
others, and the broad absorption near 4300\ang\ is weaker in SN~1994D
and SN~1998aq than in the others.  SN~2003du has strong HV\footnote{As
in Papers~I and II, PV refers to features that form at or near the
photospheric velocity and HV refers to high--velocity detached
features.} Ca~II absorptions near 3700\ang\ and 8000\ang\ (although
not nearly as strong as in the exceptional SN~2001el).  SN~1990N also
has a strong absorption near 3700\ang.  Diversity in HV Ca~II was
present also in maximum--light spectra of core--normals.

The SYNOW fitting parameters for core normals of the one--week premax
sample are in Table~2.  An example fit, to the day~$-7$ spectrum of
SN~2003du, is shown in Figure~3.  The fit is good and most of the line
identifications seem clear.  The most obvious discrepancy, the failure
of the synthetic spectrum to match the observed absorption near
3980\ang, is primarily caused by Si~II \lam4130, so we can not remove
this discrepancy without doing harm to the fit elsewhere.

The synthetic absorption produced by Si~III \lam4560 is a bit too blue
to match the observed absorption at 4400\ang\ (which is present in
most observed spectra of this paper).  A better fit could be achieved
by using C~III \lam4649 instead of Si~III.  But in a synthetic
spectrum that contains only Si~III lines the \lam4560 absorption does
occur at the right place to account for the 4400\ang\ absorption, and
Si~III \lam5743 also accounts for a weak observed absorption near
5560\ang.  In most premaximum spectra there is little or no evidence
for C~II features, so it would be surprising to see ubiquitous
evidence for the high--excitation (29.6 eV) C~III line.  Therefore,
even though neither Si~III feature fits well in the full synthetic
spectrum (the \lam5743 absorption does not even appear as a distinct
feature), we assume that these two Si~III lines usually are
responsible for the corresponding observed features.

The synthetic absorption produced by Si~II \lam5972 does not account
well for the observed absorption at 5750\ang.  This and the other two
discrepancies mentioned above are generic to our fits.  The fit could
be improved by invoking Fe~III lines with a higher excitation
potential, at or approaching 15,000~K, so that lines near 6000\ang\
contribute to the 5750\ang\ absorption [see Branch et~al. (2003) on
SN~1998aq].  But the problem with the 5750\ang\ absorption occurs in
some spectra in which we see no evidence for the stronger Fe~III lines
in the blue, so we refrain from using high--excitation Fe~III to
resolve this discrepancy.  Another way to improve the fit, in some
cases, would be to use a higher Si~II optical depth and impose a
maximum velocity on Si~II (see \S3 and \S4).

In our fits to the core normals of the one--week premax sample, the
relative weakness of the absorption near 4300\ang\ in SN~1994D and
SN~1998aq is accounted for by using lower Mg~II and Fe~III optical
depths (see Table~2).  In the same two SNe we have used C~II lines, to
allow \lam6580 to improve the fit near the peak of the Si~II \lam6355
feature.

Figure~4 shows the spectra of the three core--normals (and the two
broad--lines, to be discussed in \S3) of the early sample.  The
diversity among the core--normals of the early sample is much greater
than in the one--week premax sample, e.g., the three 6100~\AA\
absorptions have distinctly different shapes, and in SN~1994D the
absorption trough from 4600 to 5000\ang\ is deeper than in SN~1990N
and SN~2003du.  The early spectrum of SN~1994D is an example of group
crossover: the spectrum is more like that of the broad--line SN~2002bo
than like the other core--normals.  (On the other hand, the spectrum
of the broad--line SN~2002er could pass for core--normal.)

The SYNOW fitting parameters for the three core--normals (and the two
broad--lines) of the early sample are in Table~3.  The deeper
4600\ang\ to 5000\ang\ absorption trough in SN~1994D is accounted for
primarily by using stronger HV Fe~II.  

How to fit the various shapes of the 6100\ang\ absorption is an
interesting issue.  For SN~1994D we fit it in the conventional way,
with a simple PV Si~II component (a fit is shown in Paper~I).  For
SN~2003du, the feature appears to consist of two marginally resolved
components (Stanishev et~al. 2007), so we use both PV and HV Si~II.
The fit is shown in Figure~5.  (We did not need to use HV Si~II for
the one--week premax spectrum of SN~2003du.)

The unusual, nearly flat--bottomed 6100\ang\ absorption of the early
spectrum of SN~1990N has been discussed by several authors (Jeffery
et~al. 1992; Fisher et~al. 1997; Mazzali 2001). Mattila et~al. (2005)
showed that the one--week premax spectrum of SN~2001el resembled the
early spectrum of SN~1990N (see their Figure~1).  As can be seen by
comparing our Figures~2 and 4, the HV absorption of the Ca~II infrared
triplet is much deeper in SN~2001el (requiring a covering factor
approaching unity), but otherwise the two spectra, including the
6100\ang\ absorption, are quite similar.  Mattila et~al. attributed
the flat--bottomed shape of the 6100\ang\ absorption to line formation
in a shell that is geometrically thin compared to the size of the
photosphere, but this explanation is not viable.  Line formation in a
thin shell that is not detached from the photosphere does produce a
(shallow) flat--bottomed absorption [see Figure~8 of Jeffery \& Branch
(1990)], but the blue edge of the absorption is blueshifted only by an
amount that corresponds to the photospheric velocity, which is not the
case in the spectra of SN~1990N and SN~2001el.  We can nicely fit the
flat--bottomed absorptions with a single Si~II component, using a high
value of $v_e$ and imposing a maximum velocity.  However, since the
blue edge of the absorption corresponds to a high velocity comparable
to that at which the HV Ca~II and HV Fe~II are forming ($\sim20,000$
\kms), it seems more reasonable to assume that the the feature is is
an unresolved blend of PV and HV components, and fit the absorption
accordingly (see Table~2 for SN~2001el and Table~3 for SN~1990N).
Support for this interpretation is provided by a previously
unpublished spectrum of SN~1990N obtained at day~$-12$ in bad weather
by S.~Benetti and M.~Turatto (Figure~6). The absorption feature
resembles the marginally--resolved two--component absorptions of
SN~2003du (Figure~5) and SN~2005cg (Figure~13).  [see also Garavini et
  al. 2007 on 05cf]

\section{BROAD--LINES}

Figure~7 shows the spectrum of the core--normal SN~2003du (for
comparison) and the spectra of the four broad--lines of the one--week
premax sample.  In maximum--light spectra (Paper~II) broad--lines have
the same spectral features and line identifications as core--normals,
but the equivalent width of \6100\ is larger and the features are
generally broader.  The four broad--lines of Figure~7 also have these
characteristics, although as in Paper~II, the borderline case SN~1992A
differs from core--normal mainly in the depth and equivalent width of
\6100.  SN~2002er is much like SN~1992A except for a weaker absorption
near 7450\ang.  As at maximum light, SN~1984A is the most extreme
broad--line in the sample.

The SYNOW fitting parameters for broad--lines of the one--week premax
sample are listed in Table~4. As at maximum light (Paper~II), the
one--week premaximum spectra of the broad--lines can be fitted with
the same ions as used for the core--normals, but with higher values of
$v_e$ and/or $\tau$.\footnote{When inspecting fitting parameters one
should keep in mind the partial degeneracy of $v_e$ and $\tau$.  For
example, for SN~2002bo the synthetic absorption produced by Si~II
\lam6355 with $\tau = 7$ and $v_e = 2000$ \kms\ is roughly similar to
the corresponding absorption in SN~1992A with $\tau = 45$ and $v_e =
1000$ \kms.}  Figure~8 shows our fit to the day~$-7$ spectrum of the
extreme broad--line SN~1984A.  For Si~II a high value of $v_e=4000$
\kms\ is used and a maximum velocity of 27,000~\kms\ is imposed.

The spectra of the two broad--lines of the early sample are shown in
Figure~4, and fitting parameters are in Table~3.  Fe~III is not used
for the early spectra of the broad--lines, but its features could be
hidden by the strong HV Fe~II lines.  Figure~9 shows our fit to
the day~$-14$ spectrum of SN~2002bo.  In an attempt to account for the
5750\ang\ absorption with Si~II \lam5972 we have used a large optical
depth of 30 for Si~II and imposed a maximum velocity of 25,000 \kms.
A maximum velocity of 22,000 \kms\ has been imposed on S~II.

\section{COOLS}

Figure~10 shows the spectrum of the core--normal SN~2003du and the
spectra of the three cools of the one--week premax sample.  For
reasons given in Paper~II, the borderline SN~1989B was included in the
cool group even though it did not show the blue trough from about
4000\ang\ to 4300\ang\ that is characteristic of the other cools.
Here again SN~1989B has a higher \W610\ value than core--normals and
resembles SN~1986G at wavelengths longer than about 4400\ang, while
lacking the blue trough.  SN~1989B remains a borderline case.

The SYNOW fitting parameters for cools of the one--week premax sample
are listed in Table~5.  At maximum light (Paper~II) our strategy when
attempting to simultaneously account for the 5750~\AA\ and 6100~\AA\
absorptions was to use a high optical depth and impose a maximum
velocity on Si~II (as we have done in this paper for some of the
broad--lines; see \S3).  Figure~11 shows our fit to the day~$-5$
spectrum of SN~1999by, where we have used the same strategy with only
partial success; in the synthetic spectrum, the 5750~\AA\ absorption
is too weak even though the 6100~\AA\ is too strong.  We regard the
line identifications in Figure~11 to be definite, except for Sc~II,
which helps in one place but cannot be considered definite.

There are no cools in the early sample.  Only a small number of cools
have been observed at all, and because they have short rise times and
are dim relative to other SNe~Ia, no spectra have been obtained at
times as early as day~$-11$.

\section{SHALLOW--SILICONS}

Figure~12 shows the spectra of the core--normal SN~2003du and the
seven shallow--silicons of the one--week premax sample.  These were
classified as shallow--silicon on the basis of their maximum--light
spectra but they also have shallow--silicon at one--week premax.
(SN~1999ac is borderline in this respect but it is more like the
shallow--silicons than the core normals in other respects, such as in
having weak S~II absorptions.)  There is plenty of diversity, however,
extending to the extreme cases of SN~1991T and its near twin SN~1997br
which at a glance do not necessarily have Si~II at all (although we do
include it in our fits), and to SN~2005hk.  SN~2005hk is in our
one--week premax sample of shallow--silicons (with fitting parameters
in Table~6 the same as used in Chornock et~al. 2006), and SN~2002cx
was in our maximum--light sample of shallow--silicons.  But as
discussed in Paper~II and elsewhere (Li et~al. 2003; Branch
et~al. 2004b; Jha et~al. 2006; Chornock et~al. 2006; Stanishev
et~al. 2007; Phillips et~al. 2007), in spite of certain spectroscopic
similarities to SN~1991T--likes, the SN~2002cx--likes appear to be
physically distinct from other SNe~Ia.

The fitting parameters for the shallow--silicons of the one--week
premax sample are in Table~6.  The shape of the 6100\ang\ absorption
in SN~2005cg is especially interesting, because it appears to consist
of marginally resolved PV and HV components of Si~II.  Our fit is
shown in Figure~13.  Note that in Figure~12 the 6100\ang\ absorption
of SN~1999ee bears some resemblance to that of SN~2005cg.  As with the
one--week premax spectrum of SN~2001el and the early spectrum of
SN~1990N, the 6100\ang\ absorption of SN~1999ee can be fit with a
single broad PV component or an unresolved combination of a PV and an
HV feature.  For the same reasoning as in the discussion of SN~1990N
and SN~2001el and the results of Garavini et~al. (2007) and Stanishev
et~al. (2007), we use the two components.

Figure~14 shows the spectra of the core--normal SN~2003du and the
three shallow--silicons of the early sample.  Again SN~1999aa and
SN~1991T do have shallow silicon, and again SN~1999ac is borderline.
Figure~15 shows our fit to the day~$-12$ spectrum of SN~1999aa.  Like
Garavini et~al. (2004), we have resorted to C III, but the
identification is not definite.  When fitting the HV Ca~II infrared
triplet absorption, the HV Ca~II H\&K absorption comes out too deep;
this happens in some of our other fits, although less than here.  We
believe the absorption feature near 5500\ang\ to be Si~III \lam5743,
but as usual it does not appear as a distinct feature in the synthetic
spectrum.

\section{DISCUSSION}

The previous sections have been organized in terms of the four groups
of Paper~II, with all group assignments based on the appearance of
near--maximum--light spectra.  To a large extent the one--week premax
spectra of this paper retain the defining characteristics of the four
groups, i.e., group assignments based on one--week premax spectra
would be much the same.  To a lesser extent, the same is true of the
early spectra of this paper, e.g., almost all shallow--silicons are
distinct from core--normals (Figure~14).  But there is some crossover
between groups in the early sample, e.g., Figure~4 shows that the
early spectra of the core--normal SN~1994D and the broad--line
SN~2002er are not very different.  In any case, as in Paper~II we see
little or no evidence that SNe~Ia actually break up into distinct
groups, except for the SN~2002cx--likes (and possibly the cools, but
see Pastorello et~al. 2007 on SN~2004eo for evidence of a connection
between core normals and cools).

The SYNOW fits to the premaximum spectra of this paper generally are
of the same quality as the fits to the maximum--light spectra of
Paper~II.  All strong features and most weak ones can be accounted for
in a plausible way.  Most line identifications seem clear, although as
discussed in previous sections there are some ambiguities, and some of
the weak features remain unidentified.  For the most part our line
identifications in premaximum spectra are consistent with those
suggested previously on the basis of SYNOW (or SYNOW--like) synthetic
spectra, as well as synthetic spectra calculated with the
Mazzali--Lucy Monte Carlo code.  Such studies have been carried out
for SN~1990N (Jeffery et~al. 1992; Mazzali et~al. 1993; Mazzali 2001),
SN~1991T (Ruiz--Lapuente et~al. 1992; Jeffery et~al. 1992; Mazzali,
Danziger, \& Turatto 1995; Fisher et~al. 1999), SN~1997br (Hatano
et~al. 2002), SN~1998aq (Branch et~al. 2003), SN~1999aa (Garavini
et~al. 2004), SN~1999ac (Garavini et~al. 2005), SN~1999ee (Mazzali
et~al. 2005a), SN~2002bo (Benetti et~al. 2004; Stehle et~al. 2005),
SN~2003cg (Elias-Rosa et~al. 2006), and SN~2005hk (Chornock
et~al. 2006).

One of the notable differences from line identifications in some of
the previous papers is that neither in this paper nor in Paper~II do
we use PV Fe~II or Co~II (except in SN~2005hk).  In Paper~I on
SN~1994D we first used weak PV Fe~II at day ~+4, and at later epochs
it became much stronger.  This appears to be consistent with the
argument of Kasen \& Woosley (2007) that the development of extensive
line blocking by PV Fe~II (and Co~II) lines soon after maximum light
is the key to understanding the SN~Ia width--luminosity relation
(Phillips et~al. 1999).  We will return to this issue in Paper~IV, on
post--maximum SN~Ia spectra.

We do use PV Fe~III for the one--week premax sample, except in the
cools.  It is commonly said, correctly, that SN~1991T--likes are
characterized by conspicuous Fe~III features.  It is not true, though,
that the Fe~III features are particularly strong in SN~1991T--likes.
At one--week premax we use Fe~III optical depths for the
shallow--silicons that are comparable to or lower than for the
core--normals and the broad--lines.  The same was true at maximum
light (although we failed to mention it in Paper~II).  At maximum and
one--week premax, Fe~III features are conspicuous in shallow--silicons
only because there is less competition from other features.  

It is interesting to consider some of the implications of the fitting
parameters, especially the line optical depths, in the context of the
local--thermodynamic--equilbrium (LTE) calculations of Hatano
et~al. (1999a).  For example, Figure~5 of Hatano et~al. shows that in
LTE near 7500~K the Si~II optical depth can exceed unity in a
composition in which hydrogen and helium have burned to carbon and
oxygen while the mass fraction of other elements, including silicon,
is only solar.  This suggests that the presence of Si~II in the
spectrum does not necessarily require synthesized silicon.  However,
for this composition the optical depth of Fe~II is predicted to be
greater than that of Si~II, which according to us is not the case in
premaximum or near--maximum spectra.  Figure~6 of Hatano et~al. shows
that for a carbon--burned composition, which does include synthesized
silicon, the Si~II optical depth exceeds that of Fe~II by a large
factor.  Therefore, since we do not use PV Fe~II in premaximum and
maximum spectra, we expect that wherever PV Si~II is detected, it is
produced by synthesized, rather than primordial, silicon.  Another
interesting example is that for the core--normals of the one--week
premax sample, the Si~II optical depths range from four to eight (and
for the core--normals of the maximum--light sample it ranged from six
to 13).  Figure~6 of Hatano et~al., for the carbon--burned
composition, shows that a temperature difference of only 800~K is
sufficient to change the Si~II optical depth by a factor of two.  This
temperature sensitivity makes the high degree of homogeneity among the
core--normals all the more remarkable.

The HV features are intriguing.  Following the discovery of HV Ca~II
and HV Fe~II in SN~1994D (Hatano et~al. 1999b), HV Ca~II has been
recognized by many authors: see Li et~al. (2001), Thomas
et~al. (2004), and Branch et~al. (2004a) on SN~2000cx; Wang
et~al. (2003), Kasen et~al. (2003), and Mattila et~al. (2005) on
SN~2001el; Mazzali et~al. (2005a) on SN~1999ee; Gerardy et~al. (2004)
and Stanishev et~al. (2007) on SN~2003du; Quimby et~al. (2006) on
SN~2005cg; and Garavini et~al. (2007) on SN~2005cf.  In premaximum
spectra of SNe~Ia, HV Ca~II is ubiquitous (Mazzali 2005b; and this
paper), and in Paper~II and we regularly invoked HV Ca~II even at
maximum light.  The identification of HV Fe~II, which usually appears
at modest strength in a crowded spectral region, is perhaps not
definite, but we have invoked it regularly both in this paper and in
Paper~II at maximum light.  One thing that is clear, from flux spectra
(Thomas et~al. 2004; Tanaka et~al. 2006) and especially from
polarization measurements (Wang et~al. 2003; Kasen et~al. 2003), is
that in at least some cases the HV Ca~II features form in asymmetrical
structures.  The {\sl origin} of the HV features is unclear.  One
possibility is that for one reason or another they form naturally in
the ejecta (Hatano et~al. 1999b).  An alternative, favored by Gerardy
et~al. (2004) and Quimby et~al. (2006), is that they form in a shell
of restricted velocity interval and elevated density that is produced
by interaction between the ejecta and circumstellar matter (an
accretion disk, a filled Roche lobe, or a common envelope).  Our
identification of HV Fe~II, if correct, may hint that the features can
form naturally in the ejecta, because more often than not we detached
HV Fe~II at a lower velocity than HV Ca~II.  In the spectra of the
early sample, it is unlikely that the HV Fe~II features are produced
by iron resulting from decay the decay of $^{56}$Ni and $^{56}$Co,
because lines of HV Ni~II and HV Co~II also would be expected (see
Figure~9 of Hatano et~al. 1999a), but it is possible that the HV Fe~II
features are produced by iron that has been freshly synthetisized
along with $^{56}$Ni.  Another hint that the HV features may arise
naturally in the ejecta is that in some cases we see HV Si~II at about
the same velocity as HV Ca~II and HV Fe~II.  In the circumstellar
interaction scenario, the blue wing of the Si~II absorption is
predicted to be truncated rather than extended (Gerardy et~al. 2004;
Quimby et~al. 2006).  In any case, it is interesting that that HV
features, of Ca~II, Fe~II, and Si~II typically are detached not far
from 20,000 \kms.  The density and/or composition structure of most
SNe~Ia seems to have something special around 20,000 \kms, probably
including a density increase (Mazzali et~al. 2005a; Mazzali 2005b). 

The day~$-14$ spectrum of the broad--line SN~2002bo (Figure~8 and
Table~3) raises an issue about the definition of HV and PV features.
We did not use HV Ca~II or HV Si~II, and what we called HV Fe~II was
only mildly detached, at 22,000 \kms, from the photosphere at 20,000
\kms. Thus the absence of HV Ca~II or HV Si~II is a matter of
definition: Ca~II and Si~II are forming at 20,000 \kms, as in other
SNe~Ia, but they are not referred to as HV simply because of the high
velocity at the photosphere.

The issue of carbon also is interesting for constraining explosion
models.  As discussed in \S2, a C~III line could be used to improve
our fits for core--normals but we suspect that it would be spurious.
We have used C~III to improve the fits for several shallow--silicons
but in no case is the identification convincing.  Marion et~al (2006)
have established that C~I lines are not detected [except in cools:
Howell et~al. (2001)], but their conclusion that carbon is much less
abundant than oxygen depends strongly on their assumption of LTE at
5000~K.  A temperature of 5000~K seems too low to account for the
strengths of other lines (e.g., Fe~III) and at higher temperatures the
upper limit on the carbon--to--oxygen ratio would be much lower (see
Figure~5 of Hatano et~al. 1999a).

Lines of C~II offer the best prospects for detecting the presence of
carbon.  Evidence has been presented for the strongest optical C~II
line, \lam6580, in SN~1998aq (Branch et~al. 2003); SN~1990N (Jeffery
et~al. 1992; Mazzali 2001); SN~1999ac (Garavini et~al. 2005); and
SN~1994D and SN~1996X (Branch et~al. 2003 and Paper~I).  Thomas
et~al. (2007) presented the most convincing evidence, in SN~2006D.  A
\lam6580 absorption was clearly present at day~$-7$ and day~$-5$, and
it weakened as maximum light was approached.  In this paper we have
invoked C~II in these events and only one other: the early spectrum of
SN~2003du, but in this case C~II \lam6580 is being used mainly just to
beat down the Si~II \lam6355 emission peak and the identification is
not convincing.  Thus, even at premaximum, C~II is elusive, or, as
Thomas et~al. put it, sporadic, perhaps due to its tendency to appear
in clumps.  If the clumps are few in number and none happen to be in
front of the photosphere, no absorption features will be seen.  Even
if a clump is in front of the photosphere, if it has sufficiently high
line--of--sight velocity (about 15,000 \kms\ or more) the \lam6580
absorption may be blueshifted into the Si~II absorption and not be
detectable.  Therefore, although C~II is seen in only a fraction of
premaximum SN~Ia spectra, the ubiquitous presence of carbon clumps in
the outer layers of SNe~Ia is not excluded.  Unburned carbon at the
observed velocities is not characteristic of published
delayed--detonation models.

Another issue to be addressed on the basis of premaximum spectra is
that of the maximum detectable ejection velocities.  From Figures~2,
4, 7, 10, 12, and 14 we can see that Ca~II absorption is generally
detectable up to about 25,000 \kms\ and sometimes up to about 30,000
\kms.  The highest maximum velocity that has been imposed on Ca~II is
34,000 \kms\ for SN~2001el.  Detectable optical depth in Ca~II can
easily be produced with a solar mass fraction of Ca~II, so the maximum
velocities of synthesized Ca~II may be lower.  From the same figures
we can see that Si~II absorption is generally detectable up to 15,000
\kms\ and sometimes above 20,000 \kms.  The highest maximum velocity
that has been imposed on Si~II is 27,000 \kms\ for the one--week
premax spectrum of SN~1984A.  We have argued above that this
absorption is produced by synthesized silicon.  This is consistent
with the 11,900 \kms\ that Mazzali, Benetti, \& Hillebrandt (2007)
adopted for a lower limit to the top of the silicon--rich layer, but
the silicon--rich layer seems to generally extend to much higher
velocities.  Such high silicon velocities are not characteristic of
pure deflagration models.  The presence of carbon at intermediate
velocities and silicon at high velocities could be consistent with the
pulsating reverse detonation models of Bravo \& Garcia--Senz (2006).

Another velocity--related issue is the possibility that some of the
P--Cygni emission components are blueshifted, as they sometimes are in
SNe~II (Chugai 1988; Dessart \& Hiller 2005). From measurements of the
wavelengths of flux maxima Blondin et~al. (2006) inferred
emission--line blueshifts of up to 8000 \kms\ in premaximum SN~Ia
spectra.  The assumptions on which SYNOW is based (sharp photosphere
and resonant scattering) do not allow blueshifts of unblended
emission--line peaks, yet our fits do not in general have a problem
reproducing the wavelengths of the observed flux maxima.  In SYNOW
synthetic spectra the flux maxima are affected by line blending;
sometimes they are a bit too red, consistent with an emission--line
blueshift, but equally often they are a bit too blue.  In SNe~Ia, line
blending is a serious problem for establishing that emission--line
peaks are blueshifted.

The premaximum spectra of SNe~Ia exhibit a rich diversity, which even
within each of our four groups probably is multidimensional.  The
outer layers of SN~Ia ejecta differ in composition, density, and
temperature structure, and they have asymmetries - probably both
global shape asymmetries as well as composition clumping.
Understanding the observational diversity in terms of the physical
differences is a challenging task that will require more observations
and modeling.  Flux spectra have much to tell us, but in view of the
asymmetries, polarization spectra are a high priority.

We are grateful to Stefano Benetti and Massimo Turatto for providing
the previously unpublished spectrum of SN~1990N, and to all observers
who have provided their published spectra.  This work has been
supported by NSF grants AST-0204771 and AST-0506028, and NASA LTSA
grant NNG04GD36G.

\clearpage     

\begin{figure}
\includegraphics[width=.8\textwidth,angle=270]{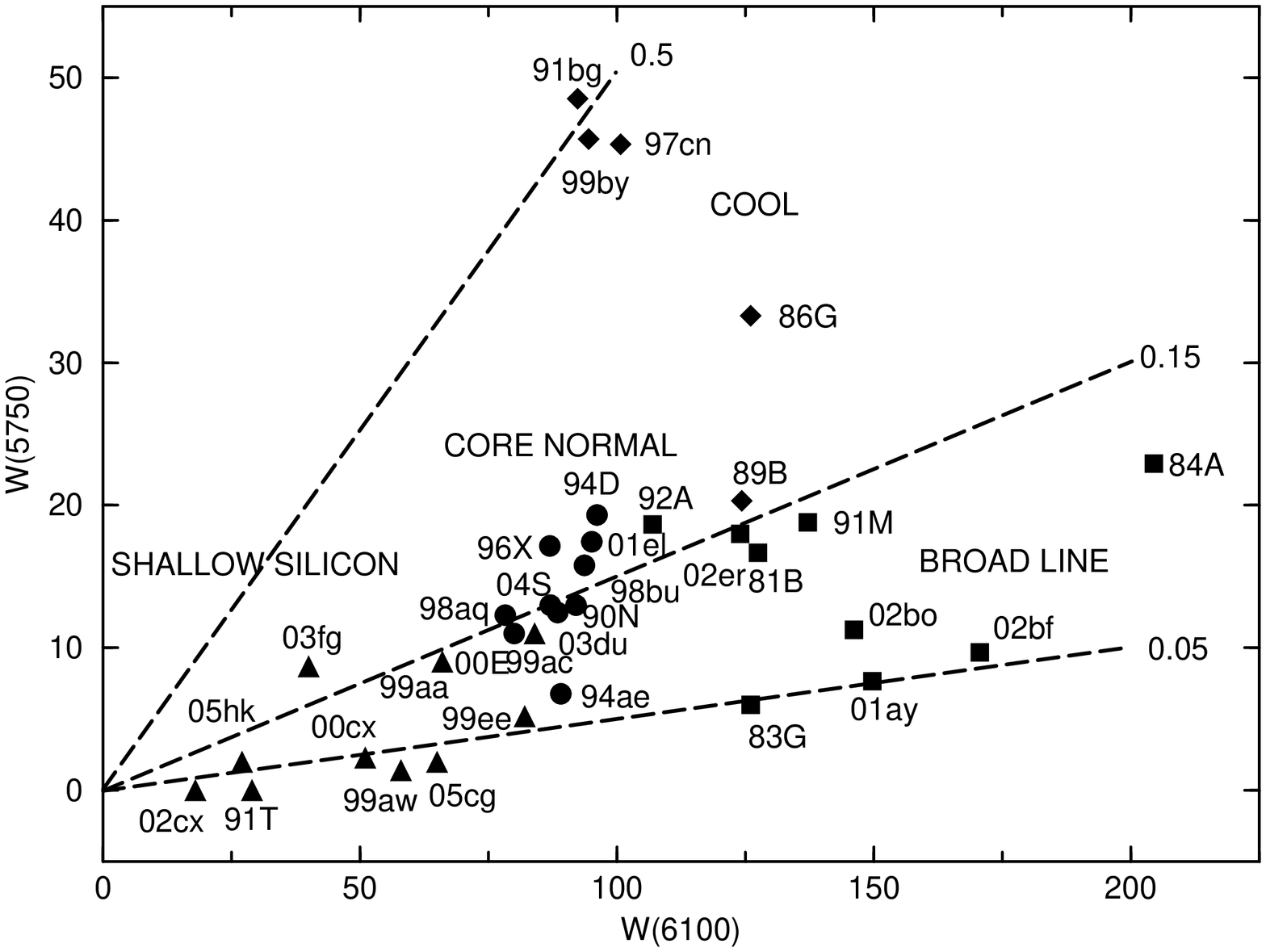}
\caption{An update of Figure~1 of Paper~II: $W(5750)$ is plotted
against $W(6100)$ for spectra obtained within three days of maximum
light.  Core--normal SNe~Ia are shown as {\sl circles}, broad--line
SNe~Ia as {\sl squares}, cool SNe~Ia as {\sl diamonds}, and
shallow--silicon SNe~Ia as {\sl triangles}. The labels of the {\sl
dashed lines} refer to the $W(5750)/W(6100)$ ratio.}
\end{figure}

\clearpage     

\begin{figure}
\includegraphics[width=.8\textwidth,angle=0]{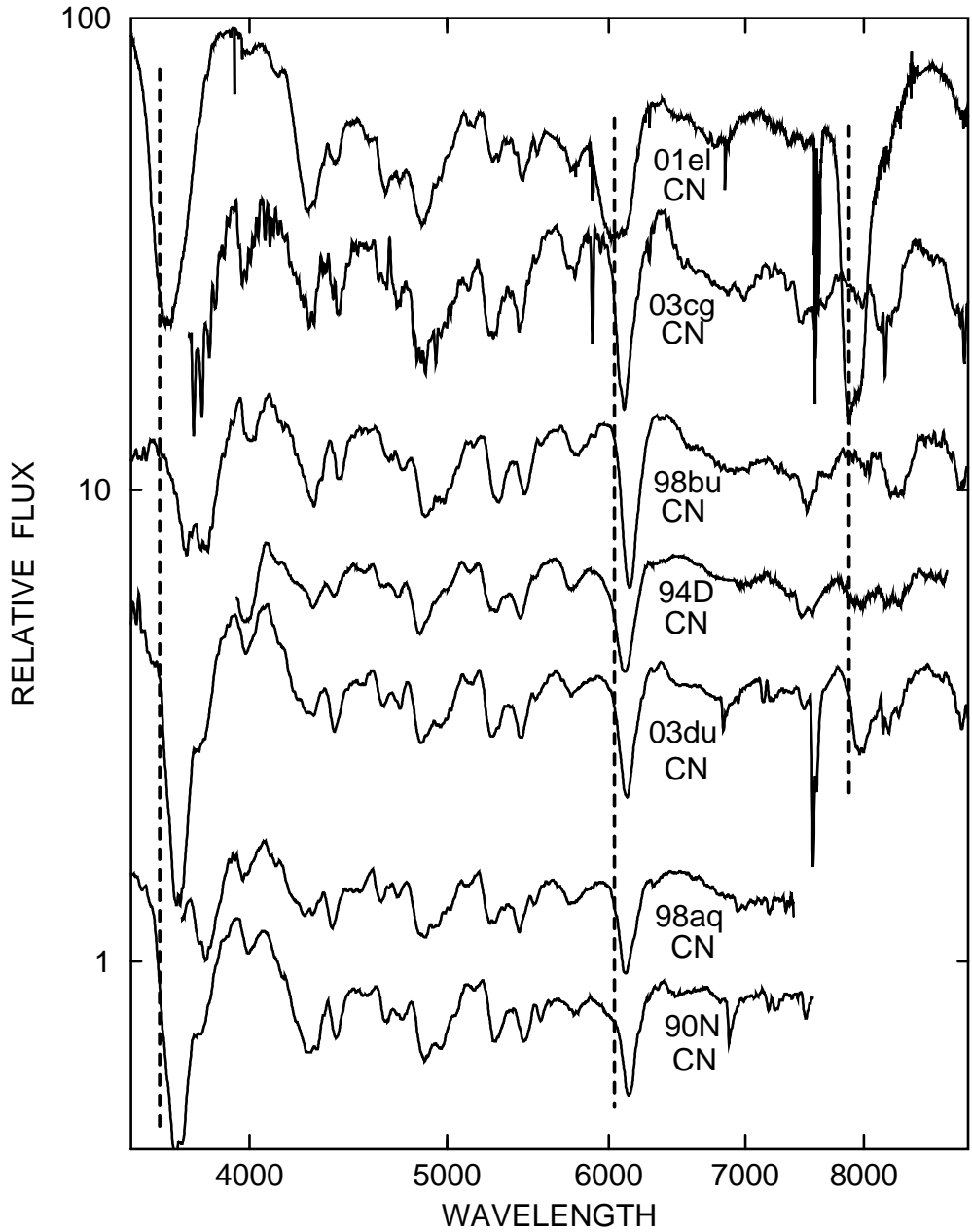}
\caption{Spectra of seven core--normals of the one--week premax
sample.  The spectra have been flattened as described in the text.
Vertical displacements are arbitrary and narrow absorptions near
7600\ang\ and 6900\ang\ are telluric.  Vertical {\sl dashed lines}
refer to Ca~II \lam3945 and \lam8579 blueshifted by 25,000 \kms\ and
Si~II \lam6355 blueshifted by 15,000 \kms.}
\end{figure}

\clearpage     

\begin{figure}
\includegraphics[width=.8\textwidth,angle=270]{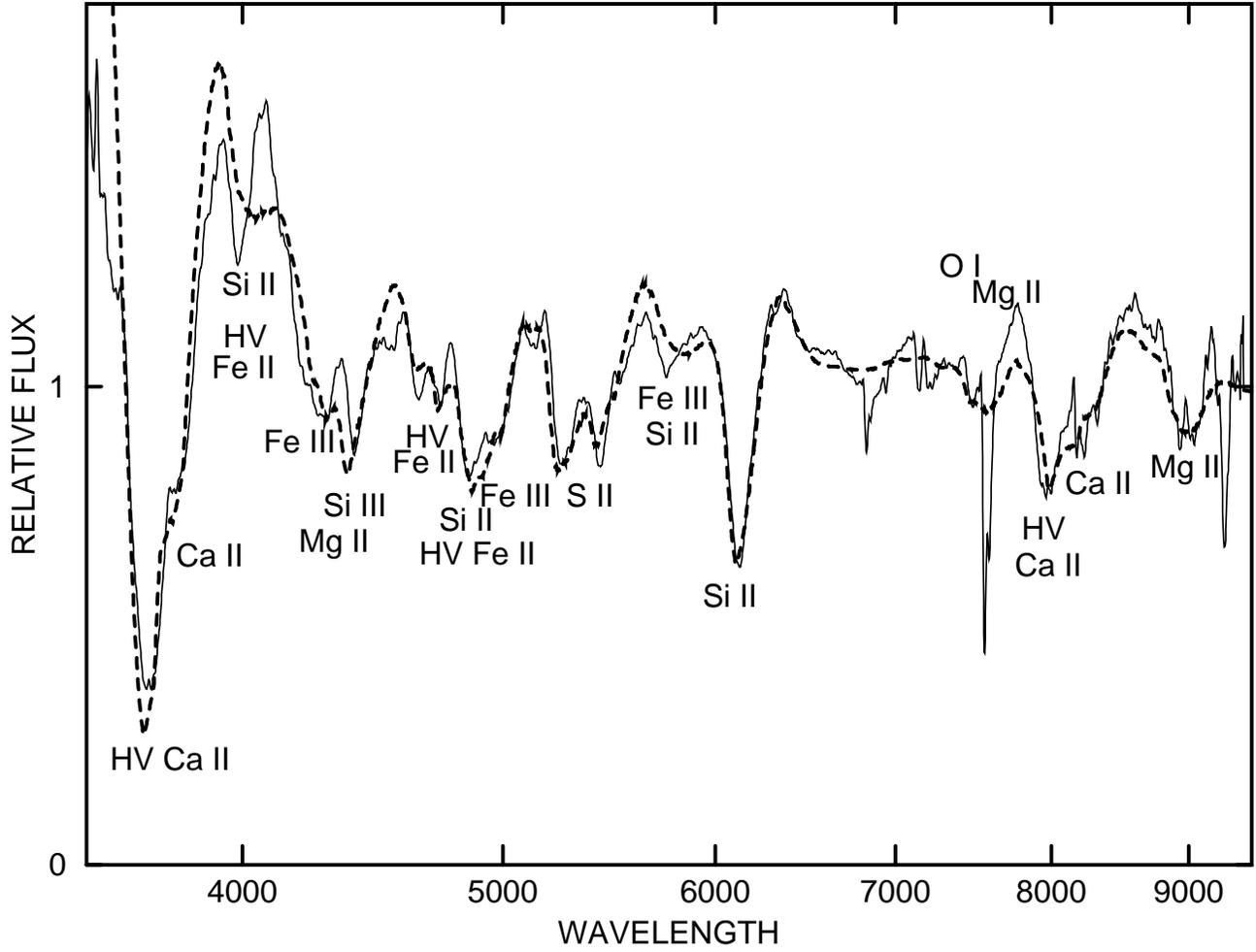}
\caption{The day~$-7$ spectrum of the core--normal SN~2003du ({\sl
  solid line}), from Anupama et~al. (2005), is compared with a
  synthetic spectrum ({\sl dashed line}).}
\end{figure}

\clearpage     

\begin{figure}
\includegraphics[width=.8\textwidth,angle=0]{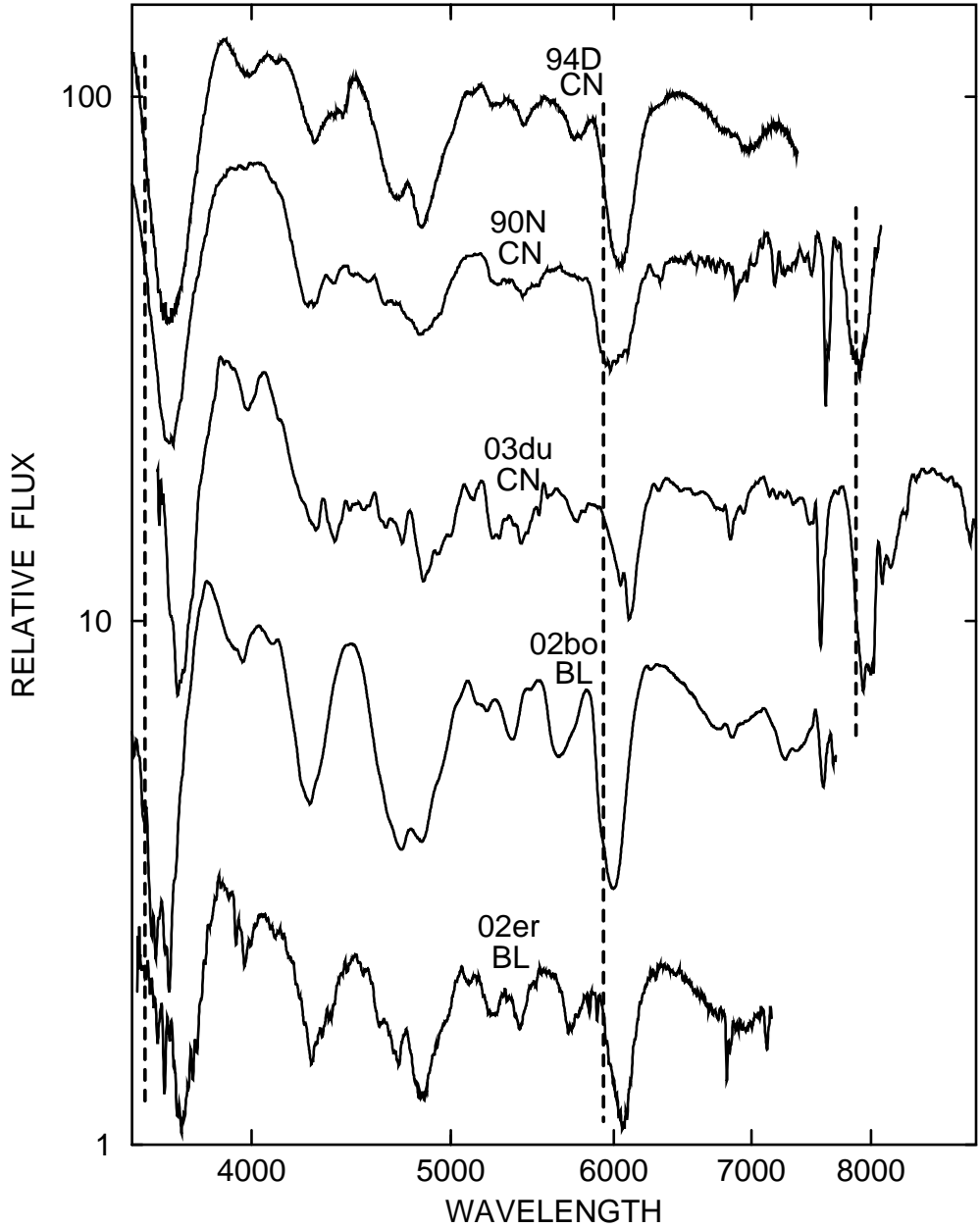}
\caption{Spectra of three core--normals and two broad--lines of
the early sample.  Vertical {\sl dashed lines}
refer to Ca~II \lam3945 blueshifted by 30,000 \kms,
Si~II \lam6355 blueshifted by 20,000 \kms, and Ca~II \lam8579 by
25,000 \kms. }
\end{figure}

\clearpage     

\begin{figure}
\includegraphics[width=.8\textwidth,angle=270]{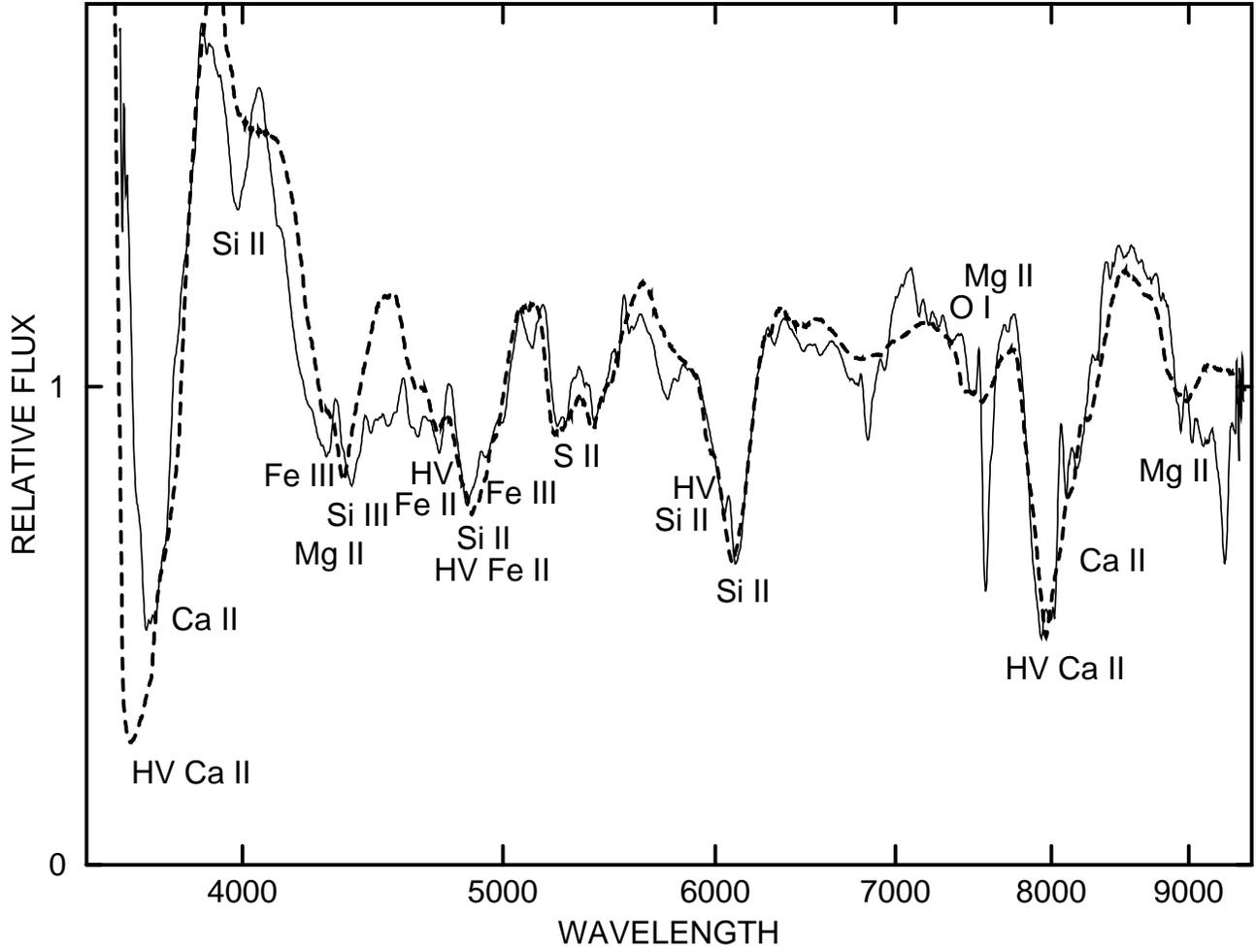}
\caption{The day~$-11$ of the core--normal SN~2003du ({\sl solid
  line}), from Stanishev et~al. (2007), is compared with a synthetic
  spectrum ({\sl dashed line}).}
\end{figure}

\clearpage     

\begin{figure}
\includegraphics[width=.8\textwidth,angle=0]{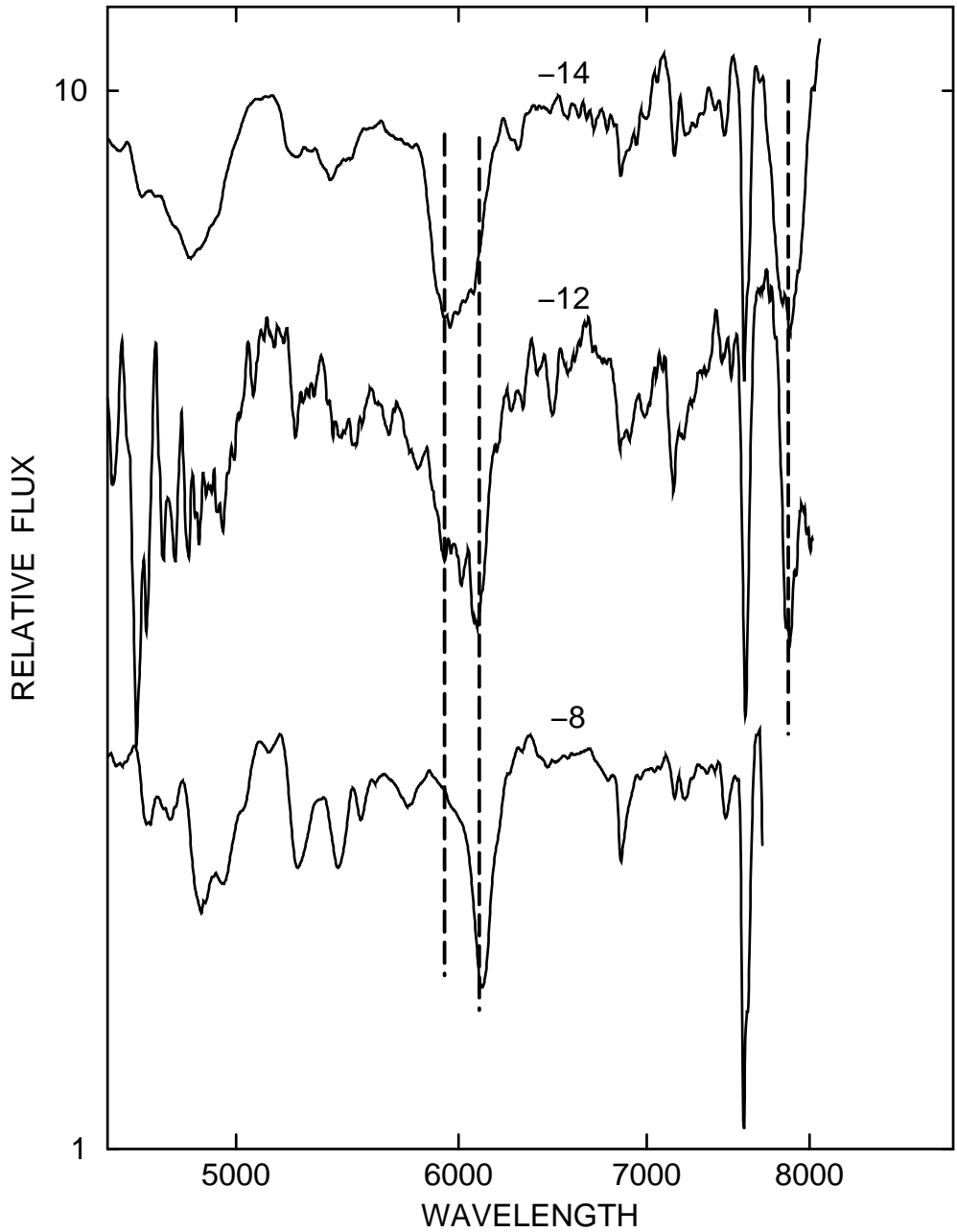}
\caption{Three spectra of SN~1990N are compared: days~$-14$ and $-8$,
  from Leibundgut et~al. (1991), and a previously unpublished spectrum
  obtained on day~$-12$ by S.~Benetti and M.~Turatto.  Vertical {\sl
  dashed lines} refer to Si~II \lam6355 blueshifted by 20,000 and
  12,000 \kms\ and Ca~II \lam8579 blueshifted by 25,000 \kms.  The
  6100\ang\ absorption in the day~$-12$ spectrum appears to be a
  marginally resolved two--component (PV + HV) feature.}
\end{figure}

\clearpage     

\begin{figure}
\includegraphics[width=.8\textwidth,angle=0]{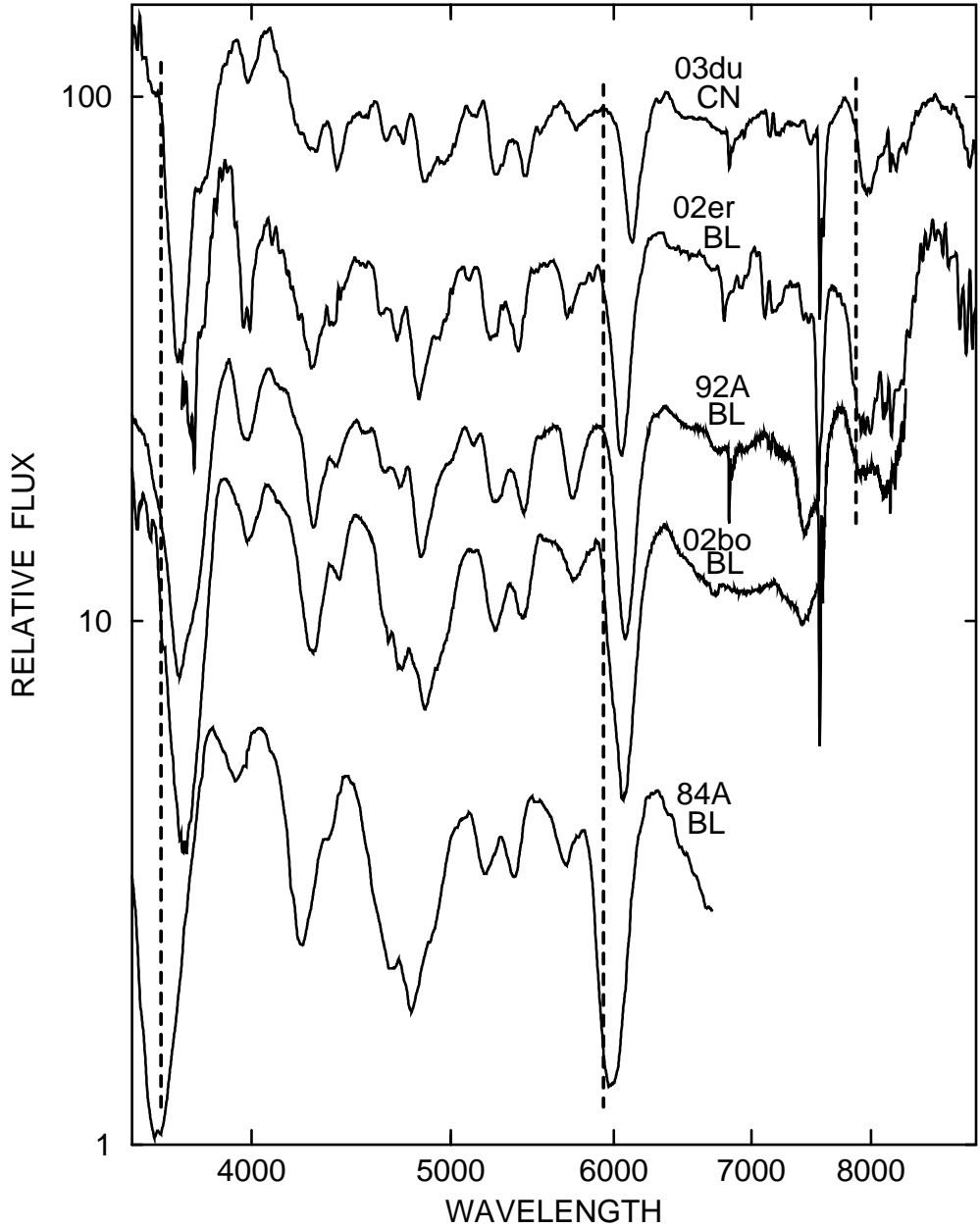}
\caption{Spectra of one core--normal and four broad--line SN~Ia of
the one--week premax sample.  Vertical {\sl dashed lines}
refer to Ca~II \lam3945 and \lam8579 blueshifted by 25,000 \kms\ and
Si~II \lam6355 blueshifted by 20,000 \kms. }
\end{figure}

\clearpage     

\begin{figure}
\includegraphics[width=.8\textwidth,angle=270]{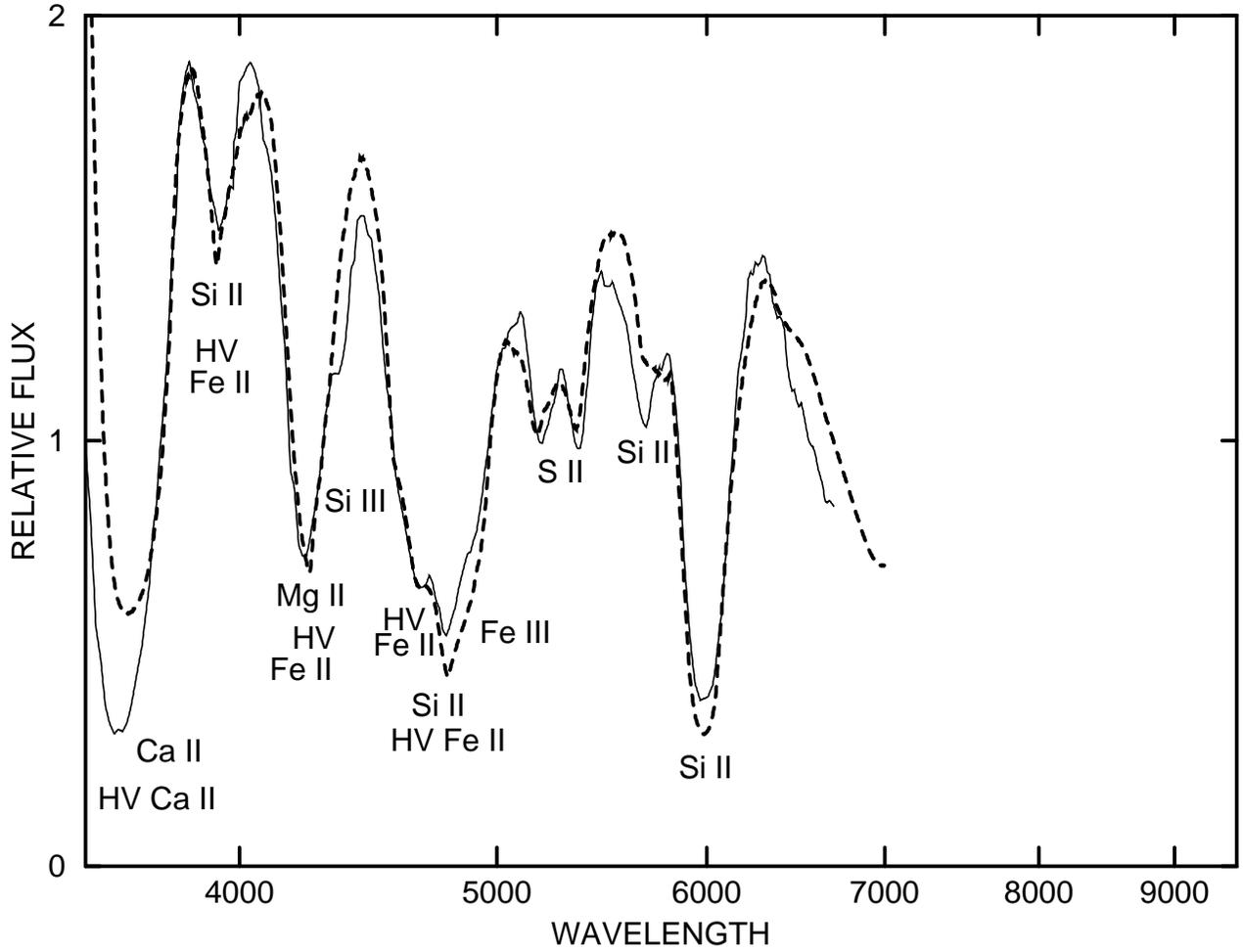}
\caption{The day~$-7$ spectrum of the extreme broad--line SN~1984A
  ({\sl solid line}), from Wegner \& McMahan (1987), is compared with
  a synthetic spectrum ({\sl dashed line}).}
\end{figure}

\clearpage     

\begin{figure}
\includegraphics[width=.8\textwidth,angle=270]{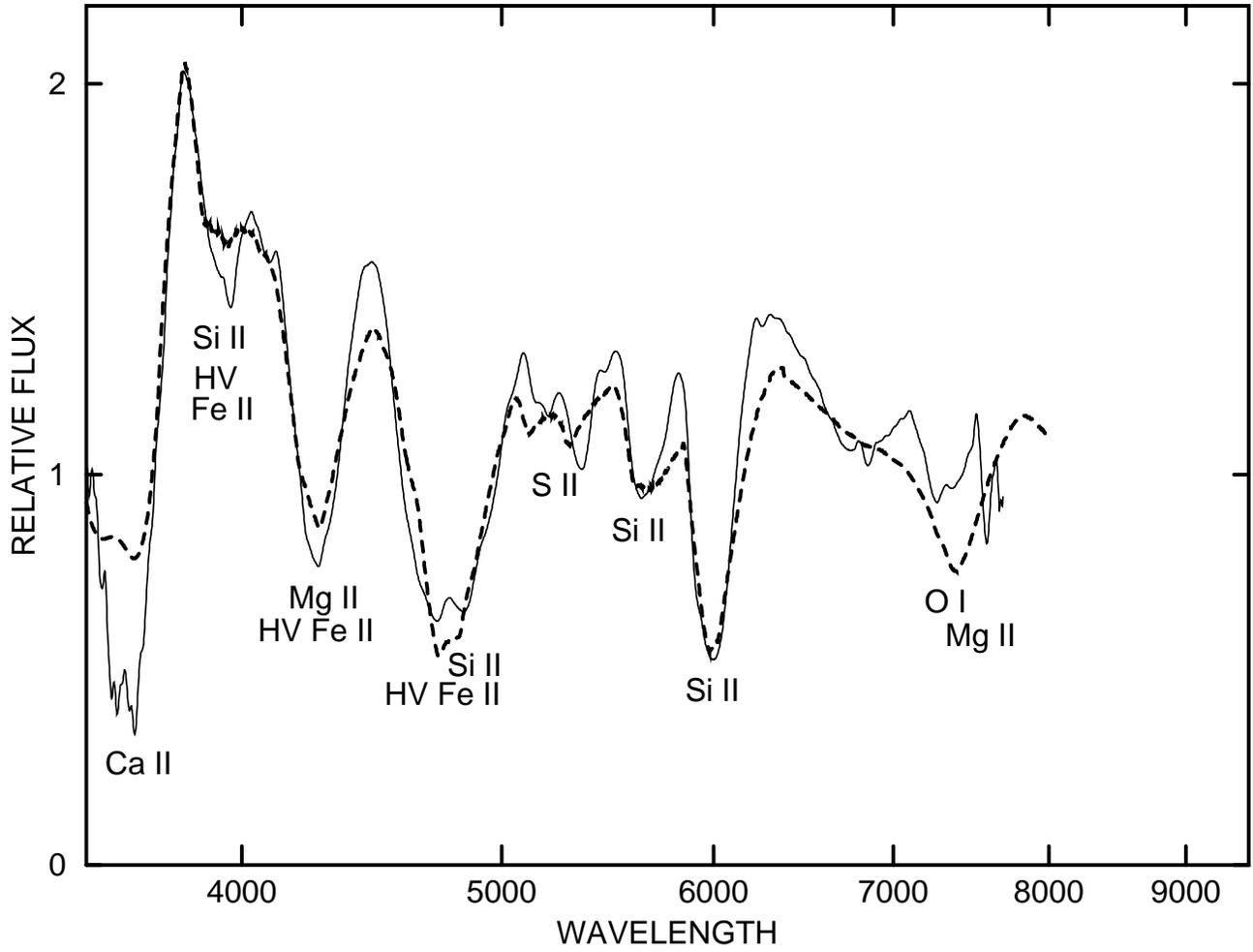}
\caption{The day~$-14$ spectrum of the broad--line SN~2002bo ({\sl
  solid line}), from Benetti et~al. (2004), is compared with a
  synthetic spectrum ({\sl dashed line}).}
\end{figure}

\clearpage     

\begin{figure}
\includegraphics[width=.8\textwidth,angle=0]{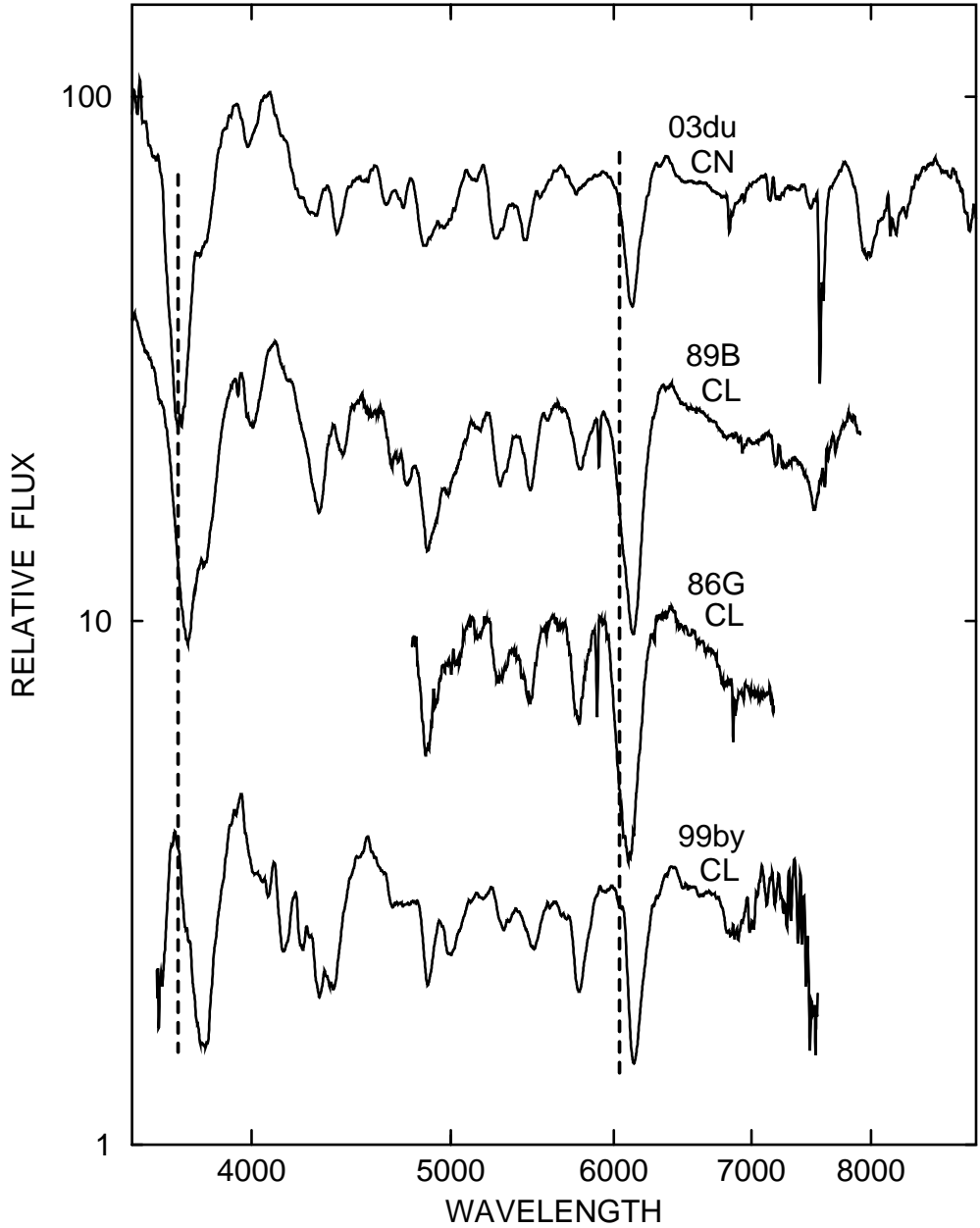}
\caption{Spectra of one core--normal and three cools of the one--week
premax sample.  Vertical {\sl dashed lines} refer to Ca~II \lam3945
blueshifted by 20,000 \kms\ and Si~II \lam6355 blueshifted by 15,000
\kms. }
\end{figure}

\clearpage     

\begin{figure}
\includegraphics[width=.8\textwidth,angle=270]{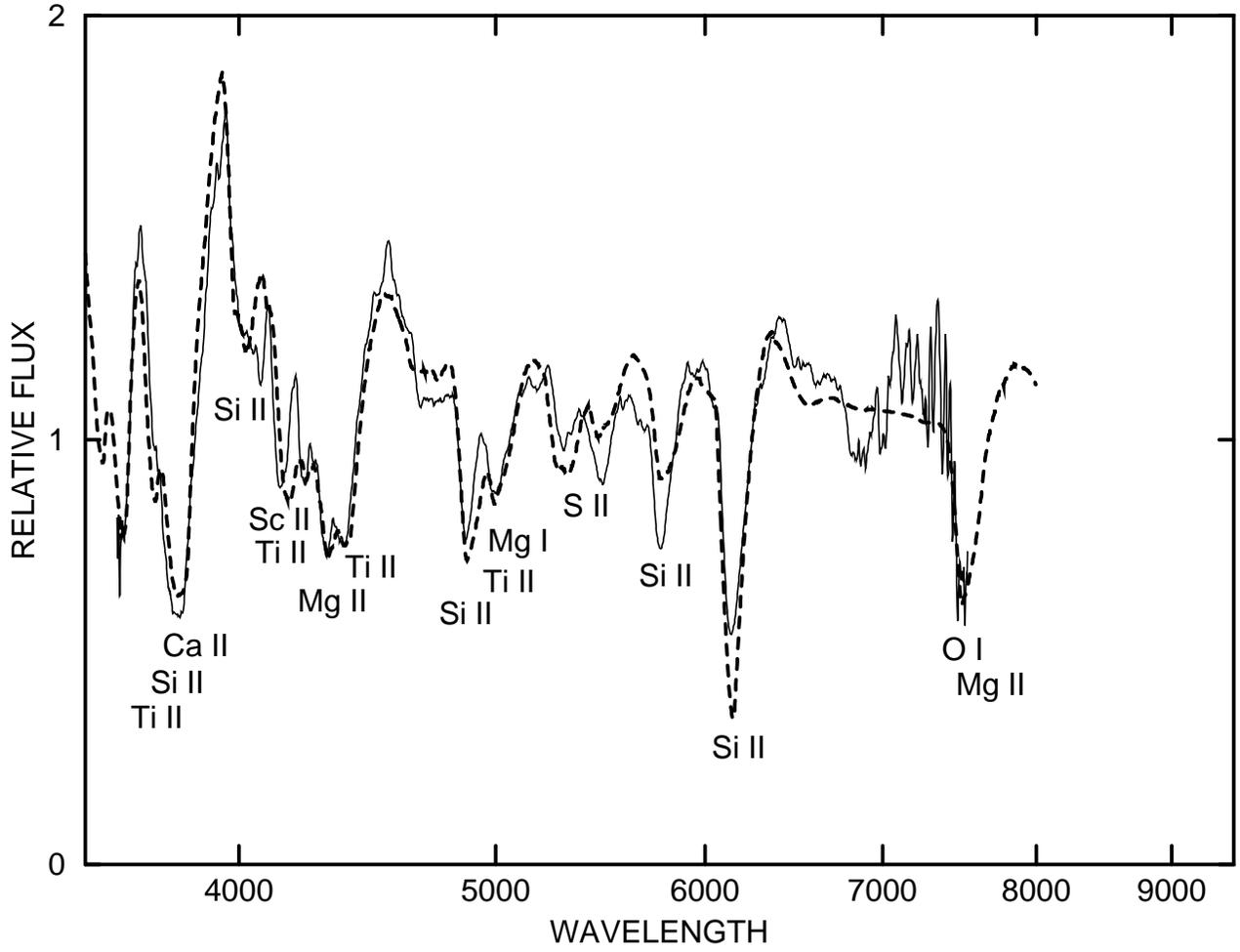}
\caption{The day~$-5$ spectrum of the cool SN~1999by ({\sl solid
  line}), from Garnavich et~al. (2004), is compared with a synthetic
  spectrum ({\sl dashed line}).}
\end{figure}

\clearpage     

\begin{figure}
\includegraphics[width=.8\textwidth,angle=0]{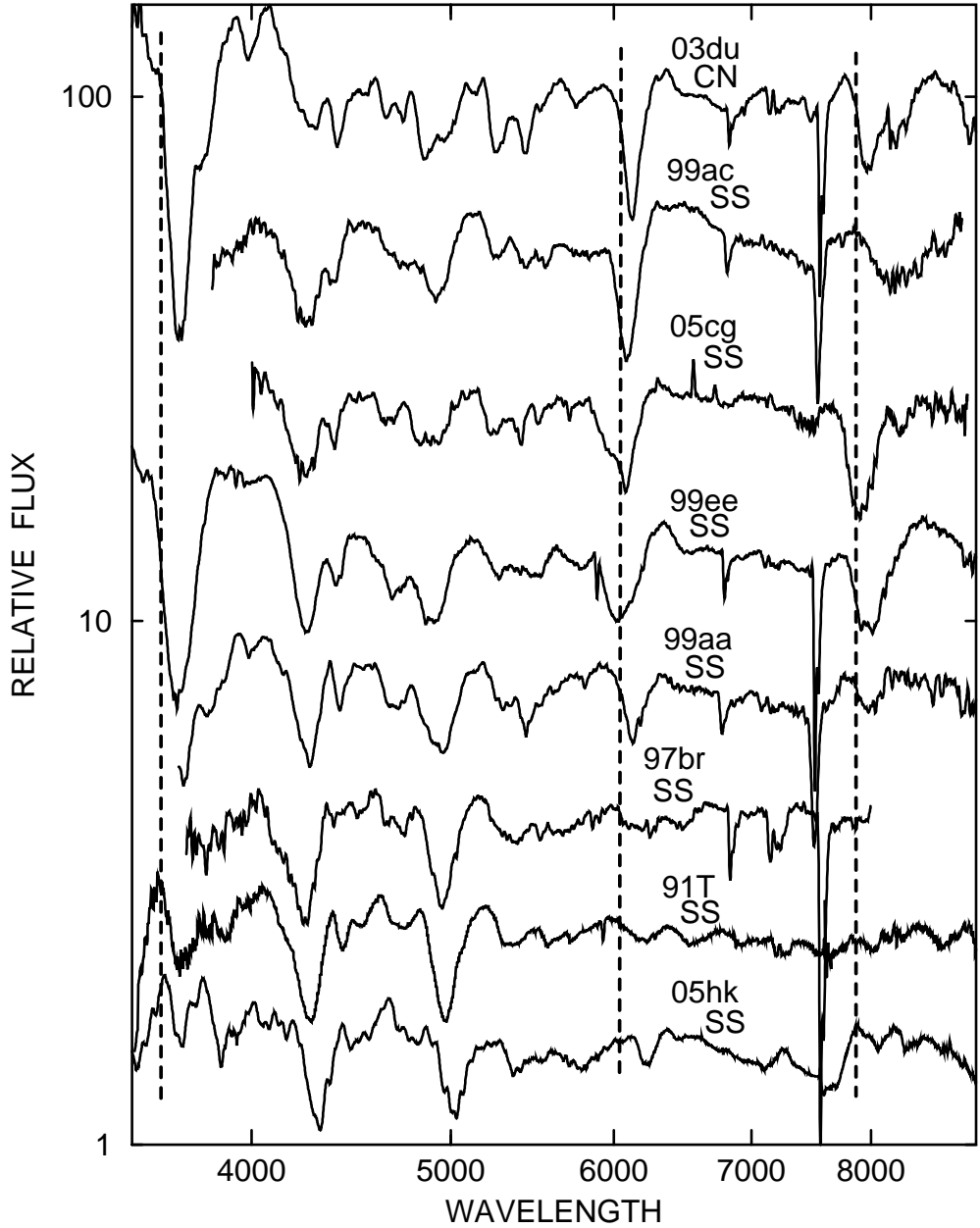}
\caption{Spectra of one core--normal and seven shallow--silicons of
the one--week premax sample.  Vertical {\sl dashed lines} refer to
Ca~II \lam3945 and \lam8579 blueshifted by 25,000 \kms\ and Si~II
\lam6355 blueshifted by 15,000 \kms. }
\end{figure}

\clearpage     

\begin{figure}
\includegraphics[width=.8\textwidth,angle=270]{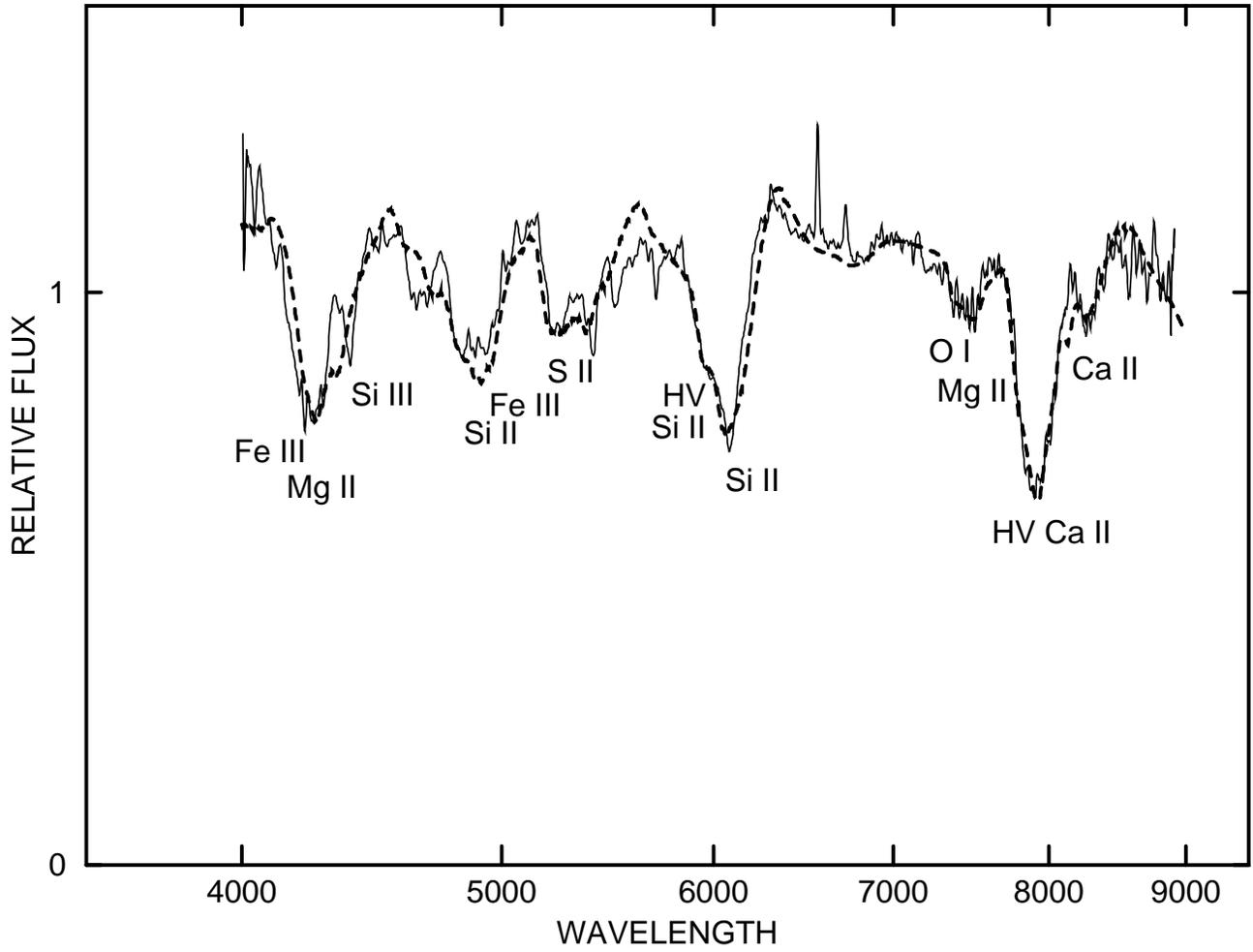}
\caption{The day~$-9$ spectrum of the shallow--silicon SN~2005cg ({\sl
  solid line}), from Quimby et~al. (2006), is compared with a
  synthetic spectrum ({\sl dashed line}).}
\end{figure}

\clearpage     

\begin{figure}
\includegraphics[width=.8\textwidth,angle=0]{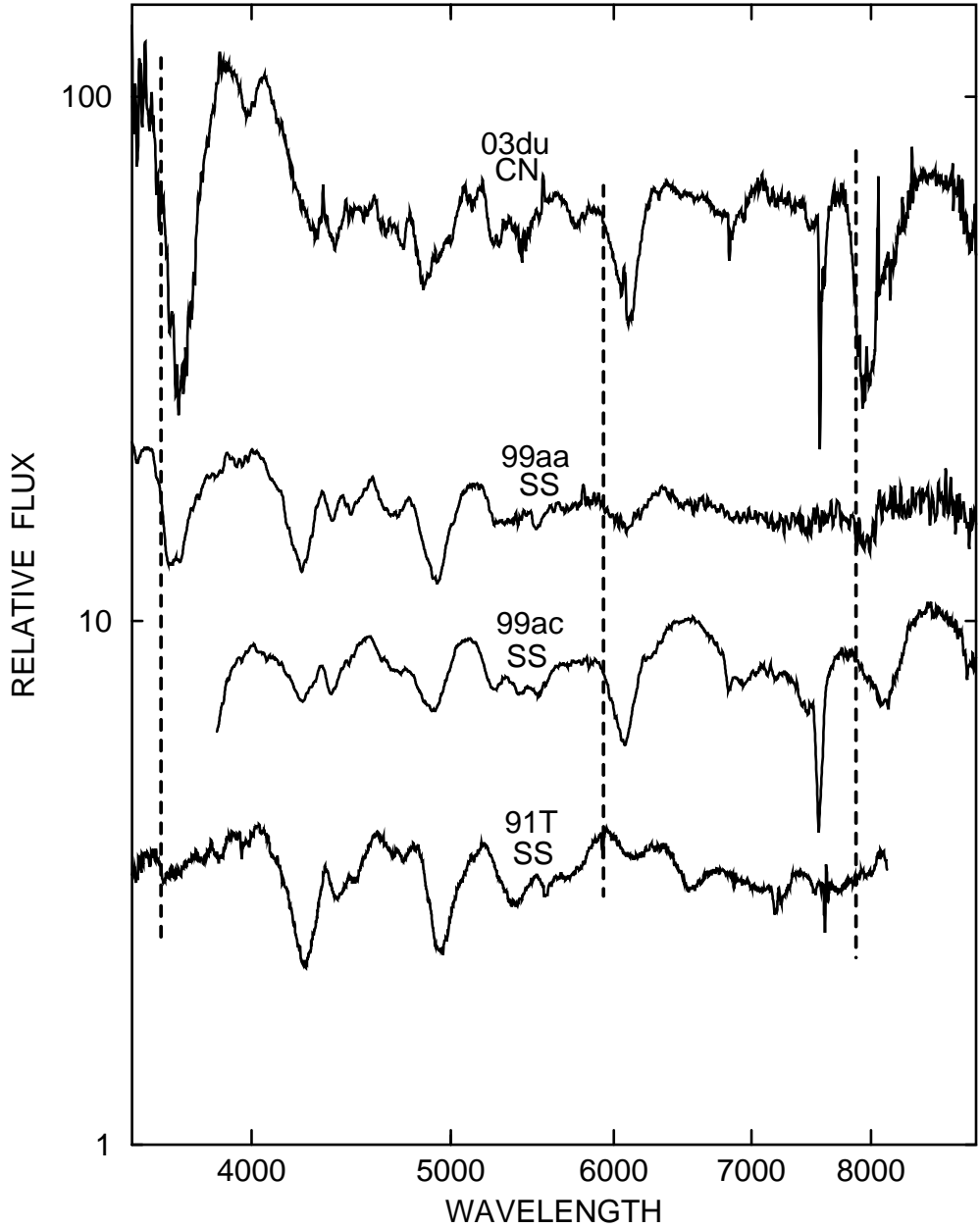}
\caption{Spectra of one core--normal and three shallow--silicons of
the early sample.  Vertical {\sl dashed lines} refer to Ca~II \lam3945
and \lam8579 blueshifted by 25,000 \kms\ and Si~II \lam6355
blueshifted by 20,000 \kms. }
\end{figure}

\clearpage     

\begin{figure}
\includegraphics[width=.8\textwidth,angle=270]{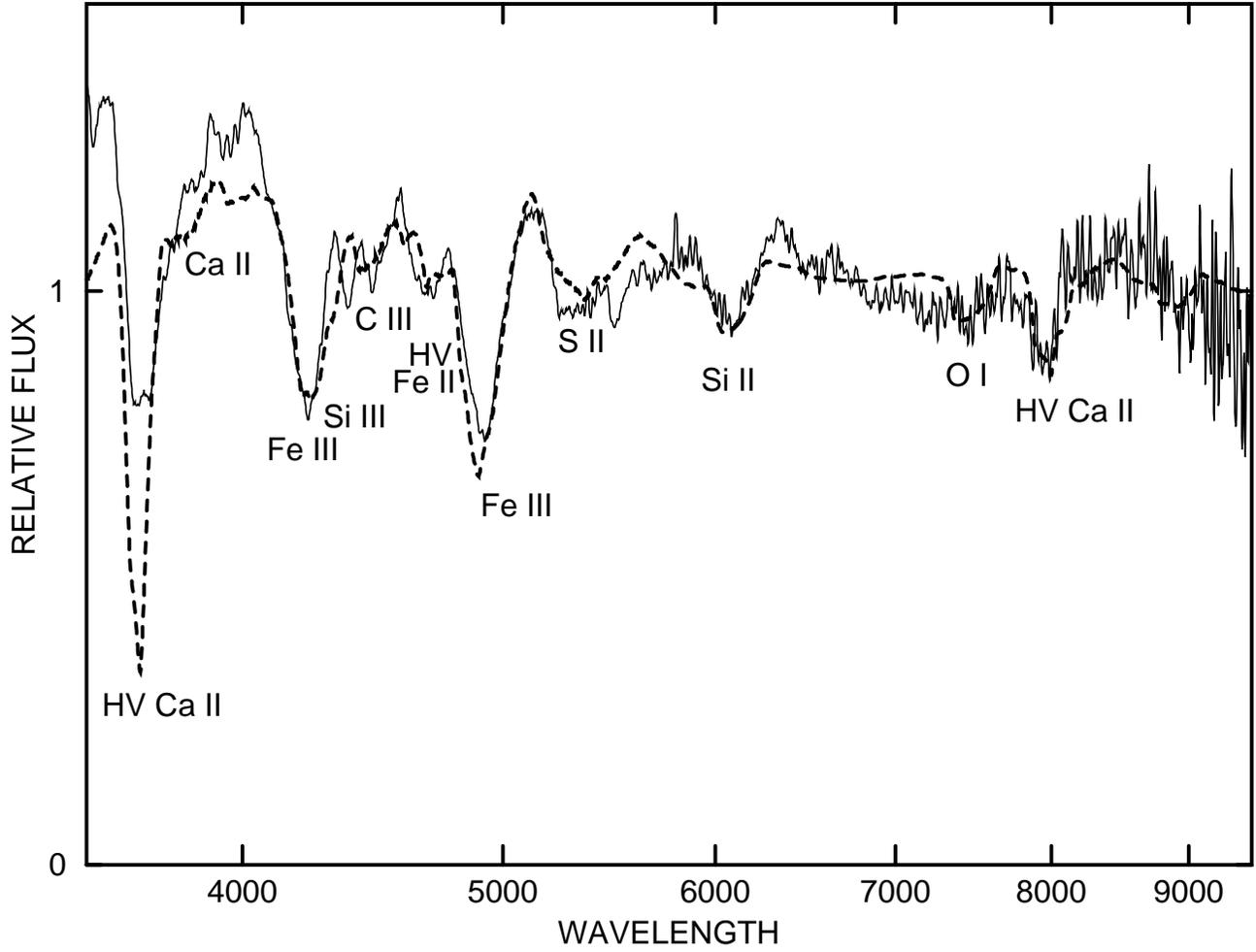}
\caption{The day~$-12$ spectrum of the shallow--silicon SN~1999aa
  ({\sl solid line}), from Garavini et~al. (2004), is compared with a
  synthetic spectrum ({\sl dashed line}).}
\end{figure}

\clearpage

\begin{deluxetable}{llll}
\tablenum{1}
\setlength{\tabcolsep}{4pt}

\tablecaption{The SN~Ia Sample}

\tablehead{\colhead{SN} & \colhead{Epochs} & \colhead{Galaxy} &
\colhead{References} \\

\colhead{} & \colhead{(days)} & \colhead{} & \colhead{} }

\startdata

1984A BL & $-$7 & NGC 4419 & Wegner \& McMahan (1987)\\

1986G CL & $-$6 & NGC 5128 & Cristiani et al. (1992)\\

1989B CL & $-$7 & NGC 3627 & Wells et al. (1994)\\

1990N CN & $-$8, $-$14 & NGC 4639 & Leibundgut et al. (1991)\\

1991T SS & $-$7, $-$11 & NGC 4527 & Phillips et~al. (1992)\\

1992A BL & $-$6 & NGC 1380 & P.~Challis, unpublished\\

1994D CN & $-$8, $-12$ & NGC 4526 & Meikle et~al. (1996)\\

1997br SS & $-7$ & ESO 576-G40 & Li et~al. (1999)\\

1998aq CN & $-$8 & NGC 3982 & Branch et al. (2003)\\

1998bu CN & $-$6 & NGC 3368 & Hernandez et~al. (2000)\\

1999aa SS & $-7$, $-$12 & NGC 4469 & Garavini et~al. (2004)\\

1999ac SS & $-9$, $-15$ & NGC 2848 & Garavini et~al. (2005) \\

1999by CL & $-$5 & NGC 2841 & Garnavich et al. (2004)\\

1999ee SS & $-$7 & IC 5179 & Hamuy et al. (2002)\\

2001el CN & $-8$ & NGC 1448 & Mattila et al. (2005)\\

2002bo BL & $-$6, $-$14 & NGC 3190 & Benetti et
al. (2004)\\

2002er BL & $-7$, $-11$ & UGC 10743 & Kotak et al. (2006)\\

2003cg CN & $-7$ & NGC 3169 & Elias--Rosa et al. (2006)\\

2003du CN & $-$7, $-11$ & NGC 1921 & Anupama et al. (2005), Stanishev
          et al. (2007)\\

2005cg SS & $-9$ & NGC 9290 & Quimby et al. (2006)\\

2005hk SS & $-5$ & UGC 272 & Chornock et al. (2006)\\

\enddata
\end{deluxetable}

\clearpage

\begin{deluxetable}{lccccccc}
\rotate
\tablenum{2}
\setlength{\tabcolsep}{4pt}

\tablecaption{Fitting Parameters for Core--Normal SNe~Ia of the
  One--Week Premax Sample}

\tablehead{\colhead{Parameter} & \colhead{SN~1990N} &
\colhead{SN~1994D} & \colhead{SN~1998aq} & \colhead{SN~1998bu} 
& \colhead{SN~2001el} & \colhead{SN~2003cg} & \colhead{SN~2003du} }

\startdata

\vphot\ (\kms) & 12,000 & 13,000 & 13,000 & 11,000 & 15,000 & 13,000 &
13,000 \\

\ve\ (\kms)& 1000 & 1000 & 1000 & 1000 & 2000 & 1000 & 1000\\

$\tau$(C~II) & \nodata & 0.8 & 0.4/14 & \nodata & \nodata & \nodata &
\nodata \\

$\tau$(O~I) & \nodata & 0.5/14 & \nodata & 0.6/12  & \nodata & 0.8 & 0.3 \\

$\tau$(Mg~II) & 1.2 & 0.5 & 0.4 & 1.5 & 0.5 & 0.7 & 0.5 \\

$\tau$(Si~II) & 4 & 5 & 5 & 8 & 2& 8 & 5 \\

$\tau$(HV Si~II) & \nodata & \nodata & \nodata & \nodata & 
0.5/20 & 8 & 5 \\

$\tau$(Si~III) & 1.1 & 0.6 & 1.8 & 1.8 & 0.2 & 1.7 & 2.2 \\

$\tau$(S~II) & 1.7 & 1.5 & 1.5 & 2 & 0.4 & 3 & 1.8\\

$\tau$(Ca~II) & 5(2) & 10 & 4 & 25 & \nodata & 25 & 6(2) \\

$\tau$(HV Ca~II) & 1.8(4)/20 & 2.5(3)/23 & \nodata &1.6(2)/20 &
20(8)/23[34] & 5/22 & 5(3)/21 \\

$\tau$(HV Fe~II) & 0.2(3)/17 & 0.25(2)/19 & 0.2(2)/19 & 0.2(2)/17 &
1/20 & 0.4/19 & 0.4/19 \\

$\tau$(Fe~III) & 1.2 & 0.5 & 0.7 & 1.5 & \nodata & 1.5 & 0.8 \\

\enddata

\tablenotetext{a}{For each ion the optical depth, $\tau$, is the
optical depth at the photosphere or detachment velocity of the ion's
reference line (ordinarily the ion's strongest line in the optical
spectrum).  When a value of \ve\ other than the default value in row~2
is used, the value (in units of 1000~\kms) is given in parentheses.
Minimum and maximum velocities (in units of 1000~\kms) are preceded by
a forward slash.  Maximum velocities are in square brackets.}

\end{deluxetable}

\begin{deluxetable}{lccccc}
\rotate
\tablenum{3}
\setlength{\tabcolsep}{4pt}

\tablecaption{Fitting Parameters for Three Core--Normal and Two
  Broad--line SNe~Ia of the Early Sample}

\tablehead{\colhead{Parameter} & \colhead{SN~1990N} &
\colhead{SN~1994D} & \colhead{SN~2003du} & \colhead{SN~2002bo}&
\colhead{SN~2002er} }

\startdata

\vphot\ (\kms) & 14,000 & 14,000 & 14,000 & 20,000 & 15,000 \\

\ve\ (\kms)& 2000 & 2000 & 1000 & 2000 & 2000 \\

$\tau$(C~II) & 0.5 & 0.6 & 0.6 & \nodata & \nodata \\

$\tau$(O~I) & \nodata & 0.5 & 0.3/15 & 0.2(3)/21 & \nodata \\

$\tau$(Mg~II) & \nodata & \nodata & 0.6 & 1 & 1 \\

$\tau$(Si~II) & 1.6 & 8 & 8 & 30[25] & 6[20]\\

$\tau$(HV Si~II) & 0.7(3)/20 & \nodata & 0.5(2)/20 & \nodata & \nodata \\

$\tau$(Si~III) & 0.3 & 0.7 & 2.5 & 0.6 & 0.2\\

$\tau$(S~II) & 0.5 & 0.6 & 1.8 & 1[22] & 1.5[16]\\

$\tau$(Ca~II) & \nodata & 200 & 12(2) & 20(10) & 10\\

$\tau$(HV Ca~II) & 8(6)/24[30] & 200(6)/23 & 12(5)/22 & \nodata & 100(6)/23\\

$\tau$(HV Fe~II) & 0.3(3)/20 & 1.2(3)/20 & 0.3(3)/19 & 1(3)/22 & 0.7(3)/20\\

$\tau$(Fe~III) & 0.5 & \nodata & 0.8 & \nodata & \nodata\\

\enddata

\tablenotetext{a}{For column descriptions see the note to Table~2.}

\end{deluxetable}

\clearpage     

\begin{deluxetable}{lcccc}
\rotate
\tablenum{4}
\setlength{\tabcolsep}{4pt}

\tablecaption{Fitting Parameters for Broad--Line SNe~Ia of the
  One--Week Premax Sample}

\tablehead{\colhead{Parameter} & \colhead{SN~1984A} &
\colhead{SN~1992A} & \colhead{SN~2002bo} &\colhead{SN~2002er} }

\startdata

\vphot\ (\kms) & 16,000 & 14,000 & 14,000 &16,000\\

\ve\ (\kms)& 2000 & 1000 & 2000 &1000\\

$\tau$(O~I) & \nodata & 1.5/16 & 0.15(3)/15 &0.1 \\

$\tau$(Mg~II) & 3 & 1.2(2) & 0.8  &1.7(2)\\

$\tau$(Si~II) & 10(4)/[27] & 45 & 7 & 35\\

$\tau$(Si~III) & 0.6 & 0.5 & 0.5 &0.7 \\

$\tau$(S~II) & 2.3 & 2.5 & 1.7 &3.5\\

$\tau$(Ca~II) & 20(6) & 50 & 120 &150\\

$\tau$(HV Ca~II) & 20(6)/21 &2(20)/24[28] & 5(3)/21 & 8(20)/24[29]\\

$\tau$(HV Fe~II) & 1.7(3)/24 & 0.8/19 & 1(3)/19 & 1.5/20\\

$\tau$(Fe~III) & 1.4 & 0.5 & 1 &0.4  \\

\enddata
\tablenotetext{a}{For column descriptions see the note to Table~2.}
\end{deluxetable}

\clearpage     

\begin{deluxetable}{lccc}
\rotate
\tablenum{5}
\setlength{\tabcolsep}{4pt}

\tablecaption{Fitting Parameters for Cool SNe~Ia of the One--Week
  Premax Sample}

\tablehead{\colhead{ } & \colhead{SN~1989B} & \colhead{SN~1986G} &
\colhead{SN~1999by} }

\startdata

\vphot\ (\kms) & 12,000 & 12,000 & 11,000  \\

\ve\ (\kms)& 1000 & 1000 & 1000  \\

$\tau$(O~I) & 1/13 & \nodata & 4  \\

$\tau$(Mg~I) & \nodata & 2.5 & 1.8 \\

$\tau$(Mg~II) & 1/13 &  \nodata & 2 \\

$\tau$(Si~II) & 20 & 40(2)/[19] & 60/[14]\\

$\tau$(Si~III) & 0.7 & \nodata & \nodata\\

$\tau$(S~II) & 2 & 2 & 1\\

$\tau$(Ca~II) & 10(2) & \nodata & 40  \\

$\tau$(HV Ca~II) & 1(2)/20 & \nodata & \nodata \\

$\tau$(Sc~II) & \nodata & \nodata & 1.5  \\

$\tau$(Ti~II) & \nodata & \nodata & 1.5  \\

$\tau$(HV Fe~II) & 1/17 & \nodata & \nodata \\

$\tau$(Fe~III) & 1.5 & \nodata & \nodata \\

\enddata
\tablenotetext{a}{For column descriptions see the note to Table~2.}
\end{deluxetable}

\clearpage     

\begin{deluxetable}{lccccccc}
\rotate
\tablenum{6}
\setlength{\tabcolsep}{4pt}

\tablecaption{Fitting Parameters for Shallow--Silicon SNe~Ia of the
  One--Week Premax Sample}

\tablehead{\colhead{ } & \colhead{SN~1999ee} & \colhead{SN~1991T} &
\colhead{SN~1997br} & \colhead{SN2005hk} & \colhead{SN~1999aa} &
\colhead{SN~1999ac} & \colhead{SN~2005cg}}

\startdata

\vphot\ (\kms)& 13,000 &  10,000 & 12,000 & 5000 & 11,000 & 13,000 & 15,000\\

\ve\ (\kms)& 2000 &  1000 & 1000 & 2000 & 2000 &2000 & 1000\\

$\tau$(C~III) & \nodata & 0.3 & \nodata & 0.3 & \nodata &
\nodata &\nodata\\

$\tau$(O~I) & 0.1 & 0.1 & \nodata & 0.08/6 & 0.1/12 & 0.1/14 &0.5/17 \\

$\tau$(Mg~II) & 0.4(3) & 0.3 & \nodata & 0.35/6 & 0.3(3) & 0.4(3)/14 &0.3(4)\\

$\tau$(Si~II) & 0.9 & 0.2 & 0.2 & 0.2 & 0.6 & 1.8 &4 \\

$\tau$(HV Si~II) & 0.4/19 & \nodata & \nodata & \nodata & \nodata &
\nodata & 0.45(2)/21\\

$\tau$(Si~III) & 0.5 & 1.3 & 0.6 & 0.3 & 0.5 & 0.9 &0.8\\

$\tau$(S~II) & 0.4 & \nodata & \nodata & 0.25  & 0.5 & 0.4 &1\\

$\tau$(Ca~II) & 5 & 1 & 0.3 & 0.8 & 1.5(3) & 4(3) &15\\

$\tau$(HV Ca~II) & 4(9)/20[28] & 0.6(3)/18 & 0.2(2)/18 & \nodata &
1.3(9)/21[28] & 1.5(6)/19[23] &5(20)/23[32]\\

$\tau$(Ti~II) & \nodata & \nodata & \nodata & 0.04 &  \nodata &
\nodata &\nodata\\

$\tau$(HV Fe~II) & 0.5/22 & \nodata & \nodata & \nodata & 0.2/21 &
0.3/19 &\nodata\\

$\tau$(Fe~III) & 0.6 & 0.7(2) & 0.8(2) & 0.7 & 0.6 & 0.5 &0.8\\

$\tau$(Co~II) & \nodata & \nodata & \nodata & 0.2 & \nodata &
\nodata &\nodata\\

$\tau$(Ni~II) & \nodata & \nodata & \nodata & 0.1 & \nodata &
\nodata &\nodata\\

$\tau$(Ni~III) & \nodata & 0.1(2) & 0.15 & 0.1 &  \nodata &\nodata &\nodata\\
\enddata
\tablenotetext{a}{For column descriptions see the note to Table~2.}
\end{deluxetable}


\clearpage     

\begin{deluxetable}{lcccc}
\rotate
\tablenum{7}
\setlength{\tabcolsep}{4pt}

\tablecaption{Fitting Parameters for Shallow--Silicon SNe~Ia of the
Early Sample}

\tablehead{\colhead{Parameter}  &
\colhead{SN~1999aa} & \colhead{SN~1999ac} & \colhead{SN~1991T} }

\startdata

\vphot\ (\kms)  & 16,000 & 17,000 & 14,000  \\

\ve\ (\kms) & 1000 & 1000 & 1000  \\

$\tau$(C~II)  & \nodata & 0.2/19 & \nodata  \\

$\tau$(C~III)  & 0.7 & \nodata & 0.5  \\

$\tau$(O~I)  & 0.3/17 & 1 & \nodata \\

$\tau$(Si~II) & 0.4/17 & 5 & 0.2 \\

$\tau$(Si~III)  & 0.6 & 2.5 & 2 \\

$\tau$(S~II)  & 0.5 & 1.2 & 0.5 \\

$\tau$(Ca~II)  & 1 &  4(2) & 1 \\

$\tau$(HV Ca~II)  & 7/25 & 1.5(5)/20 & \nodata \\

$\tau$(HV Fe~II)  & 0.5/22 & \nodata & \nodata \\

$\tau$(Fe~III)  & 2 & 1.5 & 2 \\

\enddata
\tablenotetext{a}{For column descriptions see the note to Table~2.}
\end{deluxetable}

\end{document}